\documentclass[12pt]{article}
\newlength{\abstractwidth}
\renewcommand{\thefootnote}{\fnsymbol{footnote}}
\renewcommand{\thanks}[1]{\footnote{#1}}
\newcommand{\starttext}{
\setcounter{footnote}{0}
\renewcommand{\thefootnote}{\arabic{footnote}}}

\newcommand{\bea}{\begin{eqnarray}}
\newcommand{\eea}{\end{eqnarray}}
\newcommand{\<}{\langle}
\renewcommand{\>}{\rangle}

\def\R{{\cal R}}

\def\Im{{\rm Im}}
\def\tr{{\rm tr}}

\def\p{\partial}

\def\tet{\vartheta}
\def\chiz{{\chi _{\bar z}{} ^+}}
\def\chiw{{\chi _{\bar w}{} ^+}}

\def\chiv{{\chi _{\bar v}{} ^+}}

\def\no{\nonumber}

\begin{document}
\starttext
\baselineskip=16pt
\setcounter{footnote}{0}

\begin{flushright}
UCLA/05/TEP/34 \\
Columbia/Math/05 \\
October 2005 \\
\end{flushright}

\bigskip

\begin{center} {\Large\bf COMPLEX GEOMETRY AND}\\
\bigskip
{\Large\bf SUPERGEOMETRY}
\footnote{Lecture at {\it ``Current Developments in Mathematics"},
November 2005, Cambridge, MA.
Research supported in part by National Science Foundation
grants PHY-01-40151, DMS-02-45371,
PHY-0456200, and by the KITP (Santa Barbara)
under NSF grant PHY-99-07949.}

\bigskip
\medskip

{\large Eric D'Hoker$^a$ and D.H. Phong$^b$}

\bigskip
$^a$ \sl Department of Physics and Astronomy \\
\sl University of California, Los Angeles, CA 90095, USA\\
$^b$ \sl Department of Mathematics\\
\sl Columbia University, New York, NY 10027, USA

\end{center}

\bigskip\bigskip

\begin{abstract}

Complex geometry and supergeometry are closely entertwined in superstring
perturbation theory, since perturbative superstring amplitudes are
formulated in terms of supergeometry, and yet should reduce to
integrals of holomorphic forms on the moduli space of punctured Riemann surfaces.
The presence of supermoduli has been a major obstacle for a long time
in carrying out this program.
Recently, this obstacle has been overcome at genus 2, which is the
first loop order where it appears in all amplitudes. An important
ingredient is a better understanding of the relation
between geometry and supergeometry,
and between holomorphicity and superholomorphicity. This talk provides
a survey of these developments and a brief discussion of the
directions for further investigation.

\end{abstract}

\vfill\eject

\tableofcontents

\newpage

\baselineskip=15pt
\setcounter{equation}{0}
\setcounter{footnote}{0}

\section{Introduction}
\setcounter{equation}{0}

String theory is a theory of random surfaces.
Perturbative scattering amplitudes of string states are sums
over the fluctuating worldsheets spanned by evolving strings.
Conformal invariance reduces these sums to sums over
only conformally distinct worldsheets. Thus, perturbatively,
string scattering amplitudes should be given by series
of integrals over the moduli space
${\cal M}_h$ of Riemann surfaces of genus $h\geq 0$.

\medskip
An early major success of superstring theory was
the explicit one-loop ($h=1$) amplitudes obtained
by Green and Schwarz \cite{gs82} for the superstring and by
Gross et al.\cite{Gross:1985fr} for the heterotic string.
However, the general loop order $h$ has remained intractable
to this day. This is due to a fundamental geometric difficulty
beginning at $h=2$, which is the occurrence of $2h-2$ odd {\it supermoduli}
inherent to the Neveu-Schwarz-Ramond formulation of the superstring
\cite{fms, dp86}.

\medskip

In the NSR formulation, the sums over fluctuating worldsheets for the
superstring are realized by integrating over all
{\it supergeometries} $(g_{mn},\chi_m{}^\alpha)$
instead of over all {\it geometries} $g_{mn}$, where $g_{mn}$ are metrics
on a fixed smooth surface $\Sigma$ and $\chi_m{}^\alpha$ are gravitino
fields on $\Sigma$. The standard Faddeev-Popov gauge fixing procedure reduces these sums
to integrals over the supermoduli space $s{\cal M}_h$ of inequivalent supergeometries
instead of integrals over the moduli space ${\cal M}_h$ of inequivalent geometries. The space
$s{\cal M}_h$ is a $(3h-3|2h-2)$ superspace, and
the $2h-2$ odd supermoduli have to be integrated out in
order to arrive at the desired integrals over ${\cal M}_h$.
This is a new step beyond the standard gauge fixing procedures of quantum field
theory. It is not made any easier by our insufficient understanding
of the interplay between local supersymmetry and the complex structures of Riemann
surfaces and their moduli space.

\medskip

Recently, however, the supermoduli problem has been overcome for
the case of genus $h=2$ and even spin structures
\cite{I,II,III,IV,V,VI}, which is the first loop order where it
appears in all amplitudes. The progress is based partly on
an improved understanding of the interplay between worldsheet supersymmetry
and complex structures. In particular, at genus $h=2$
and even spin structures, we have now:

\medskip

$\bullet$ A gauge-fixing procedure which reduces the sums over
fluctuating worldsheets in superstring theory to well-defined
integrals over the moduli space ${\cal M}_2$ of Riemann surfaces of
genus $h=2$, in \cite{I,II}.

\medskip
$\bullet$ These integrals are independent of the choice of gauge
slices \cite{II,III}. As pointed out in \cite{vv87, ambiguities},
gauge slice independence is a crucial requirement
which was not satisfied by the Ans\"atze for superstring
amplitudes proposed in the past.

\medskip

$\bullet$ Underlying this gauge slice independence is the
remarkable fact that gauge slice changes produce global forms
which are de Rham-exact in all insertion
points, point by point over moduli space \cite{V}.

\medskip

$\bullet$ The integrands of the superstring scattering amplitudes are
hermitian
pairings of holomorphic forms of maximal rank on the moduli space
of Riemann surfaces with punctures. Holomorphicity is a particularly
important property
for string theory, indispensable for example in the construction
of heterotic strings. The holomorphicity of the superstring integrand
is recovered from superholomorphicity by extracting a term which
is Dolbeault exact in one insertion point and de Rham exact in
the remaining insertion points \cite{V,VI}.

\medskip

$\bullet$ The measure on the moduli space of Riemann surfaces
for each spin structure $\delta$ has been evaluated in terms
of $\vartheta$-constants \cite{IV}.  It is given by a modular covariant form
$\Xi_6[\delta](\Omega)$ of weight $6$, which may be interesting in
its own right.

\medskip

$\bullet$ Using the above measure, the 0-, 1-, 2-, and 3-point scattering
amplitudes for massless NS states have been evaluated and
found to vanish identically, both for the type II and heterotic
superstrings \cite{VI}. These results provide a proof, from first principles
and to two-loop order, of ``non-renormalization theorems"
which had been conjectured \cite{old1}
on the basis of space-time supersymmetry.

\medskip

$\bullet$ The first non-vanishing two-loop amplitude, namely the
scattering amplitude of 4 massless NS bosons, has also been evaluated
explicitly, for the first time in a gauge slice independent formalism
for both the type II and heterotic superstrings \cite{VI}.
Its surprisingly simple form may give a clue to the 4-point function
for higher genus.

\medskip

$\bullet$ The two-loop string corrections to certain terms in the low energy
effective action for both type II and heterotic superstrings have been computed precisely \cite{VI,dgp}.
In particular, for the type II theories, the $\R^4$ correction
is absent, while for the heterotic theories, the $\tr F^4$,
$\tr F^2 \tr F^2$, $\R^2 \tr F^2$, and $\R^4$ corrections are all absent,
thus confirming predictions made on the basis of $S$-duality
in type IIB theory and space-time supersymmetry.
The non-vanishing two-loop correction to the $D^4\R^4$ term
in Type IIB theory has been matched precisely against earlier predictions
made on the basis of $S$-duality and space-time supersymmetry, by
Green and collaborators (joint work with M. Gutperle \cite{dgp}).

\medskip

$\bullet$ The issue of whether the two-loop cosmological constant
vanishes point by point on moduli space for certain $Z_2$-orbifold
models proposed by Kachru, Kumar, and Silverstein \cite{kks} has been resolved.
These are models with broken supersymmetry but vanishing one-loop
cosmological constant. There had been hope that the two-loop
cosmological constant would also vanish, but we find that it is not the case
(joint work with K. Aoki \cite{adp}).

\bigskip The goal of this lecture is to provide a brief survey of
these developments, with emphasis on the geometric aspects.
Superstring perturbation theory has received sustained attention
over the years \cite{vv87, old0, bis2, bis1, bis3, lech, iz, dp88},
and has motivated many mathematical developments (see e.g.
\cite{mhns, yau87} and references therein). The formulation of
superstring perturbation theory adopted here is the NSR formulation.
Though complicated by the presence of supermoduli at higher loop
level, and by the necessity to perform summations over spin
structures, the NSR formulation is based on the worldsheet action of
2-d supergravity whose quantization is well-understood, and on firm
ground. The Green-Schwarz formulation has the advantage of manifest
space-time supersymmetry and no need for supermoduli and spin
structures, but its systematic quantization beyond 1-loop order has
not yet been achieved, in part because of the presence of delicate
second class constraints. Perhaps more promising is the pure spinor
formulation of Berkovits \cite{berkovits}, which  circumvents the
second class constraints, and permits direct quantization. Yet, it
is unclear whether this formulation possesses an ungauge-fixed
action, as is customarily used for starting point. (See however 
\cite{Aisaka:2005vn}.)
\footnote{Since the  version of this paper was submitted for publication
in {\it Current Developments in Mathematics} in late September 2005,
there have been several advances in the pure spinor
formulation \cite{spinorrecent}.}

\bigskip

\noindent
{\bf Acknowledgements}

\smallskip

The authors are grateful to K. Aoki and M. Gutperle for collaboration
on parts of this project. They would like to thank Costas Bachas,
Gerard van der Geer, Michael Green, Sam Grushevky, Boris Pioline, Jacob Sturm,
Tomasz Taylor, Richard Wentworth, Edward Witten for useful
conversations, correspondences, and references.

\newpage
\section{Description of the Main Results}
\setcounter{equation}{0}

In this section, we provide a fuller description of the main results,
leaving a sketch of their derivation
to the next section.

\medskip
Our main goal is a systematic method for the evaluation of scattering amplitudes
at genus 2 of $N$ massless bosonic states in
superstring theory.
This can be viewed as the string analogue of the Feynman rules of quantum
field theory, with the two-loop diagram being a unique topological surface $\Sigma$,
and Feynman parameters given by moduli. We consider both
the type II superstring and the
$Spin(32)/{\bf Z}_2$ and $E_8\times E_8$ heterotic string theories.
In the type II superstring,
the massless bosonic states are the graviton multiplet, while they can also be
gauge bosons in the heterotic theories. The corresponding amplitudes are
functions of the $10$-dimensional momenta $k_i=(k_i^\mu)$ and
polarization tensors $\epsilon_i=(\epsilon_i^\mu)$
of the $N$ massless states, $0\leq\mu\leq 9$,
$1\leq i\leq N$. Henceforth we restrict to
genus $2$, so the moduli space ${\cal M}_2$ has dimension 3.

\subsection{General form of the scattering amplitudes}

We concentrate on the amplitude of gravitons in the type II
superstring, which we denote by ${\bf A}_{II}(k_i,\epsilon_i)$, the others following
by combining the holomorphic factors of ${\bf A}_{II}(k_i,\epsilon_i)$ with the
chiral correlators of gauge bosons, which can be computed directly. By
the chiral splitting theorem of \cite{dp89}, the amplitude ${\bf A}_{II}(k_i,\epsilon_i)$
is of the form
\bea
\label{A1}
{\bf A}_{II}(k_i,\epsilon_i)
=
\int dp_I^\mu \int_{{\cal M}_2\times\Sigma^N}
{\cal H}(z_i;k_i, \epsilon_i ;p_I^\mu)
\wedge \overline{{\cal H}(z_i;k_i, \epsilon_i ;p_I^\mu)}
\eea
Here we have fixed a canonical homology basis $A_I,B_I$,
$\#(A_I\cap A_J)=\#(B_I\cap B_J)=0$, $\#(A_I\cap B_J)
=\delta_{IJ}$. Let $\omega_I(z)$ be the basis of holomorphic $(1,0)$-forms
dual to the $A_I$ cycles, and set $\Omega_{IJ}=\oint_{B_I}\omega_J$.
The moduli space
${\cal M}_2$ is identified with a fundamental domain of $Sp(4,{\bf Z})$
in the Siegel domain of symmetric matrices $\Omega_{IJ}$ with
positive imaginary part.
The parameters $p_I^\mu$, $1\leq I\leq h$, $0\leq\mu\leq 9$ are
internal loop momenta.
The expression ${\cal H}(z_i;k_i,\epsilon_i;p_I^\mu)$ is a $\Lambda^{3,0}({\cal M}_2)
\otimes (\otimes_{i=1}^N\Lambda^{1,0}_{z_i}(\Sigma))$ form on
${\cal M}_2\times\Sigma^N$
which is holomorphic in both moduli and insertion points $z_j$ away from
$z_j\not=z_k$, but which are twisted by the following monodromy as a point $z_j$
is transported along a closed cycle
\bea
\label{monodromy}
{\cal H}(z_i+\delta_{ij}A_K;k_i,\epsilon_i;p_I^\mu)
&=&
{\cal H}(z_i;k_i,\epsilon_i;p_I^\mu)
\nonumber\\
{\cal H}(z_i+\delta_{ij}B_K;k_i,\epsilon_i;p_I^\mu)
&=&
{\cal H}(z_i;k_i,\epsilon_i;p_I^\mu+\delta_{IK}k_j^\mu)
\eea
The problem of evaluating the amplitude ${\bf A}_{II}$ reduces to that of determining
${\cal H}(z_i;k_i,\epsilon_i;p_I^\mu)$. The holomorphicity of the desired form
${\cal H}(z_i;k_i,\epsilon_i;p_I^\mu)$ is an essential requirement for the construction
of heterotic string theories.

\subsection{Holomorphic ${\cal H}$, chiral ${\cal B}[\delta]$, and
Dolbeault cohomology}

The following algorithm, based on a Dolbeault cohomology procedure, gives a solution to
the problem of finding ${\cal H}(z_i;k_i,\epsilon_i;p_I^\mu)$ \cite{V}.

\medskip

For each even spin structure $\delta$ on $\Sigma$,
there exists a form ${\cal B}[\delta](z_i;k_i,\epsilon_i;p_I^\mu)$ which is a correlation function
on the worldsheet $\Sigma$ and which can itself be evaluated explicitly.
We shall give the full prescription for ${\cal B}[\delta](z_i;k_i,\epsilon_i;p_I^\mu)$
in the next section, but for the moment, we stress that
${\cal B}[\delta](z_i;k_i,\epsilon_i;p_I)$ is a closed form in each $z_i$,
and that for $N\geq 1$,
${\cal B}[\delta](z_i;k_i,\epsilon_i;p_I^\mu)$
is a 1-form in each point $z_i$ which may incorporate $(0,1)$-components.
For such forms ${\cal B}[\delta](z_i;k_i,\epsilon_i;p_I^\mu)$, there is no notion of holomorphicity.
These forms arise from the chiral splitting theorem of \cite{dp89}
and are sometimes referred to as ``chiral", since they are built only
from correlations functions of chiral spinors on $\Sigma$. However, we stress that
they are in general not holomorphic in $z_i$.

\medskip

$\bullet$ Consider first the $N$-point function with $N=0$, which corresponds to
the cosmological constant. Then there are no insertion points $z_i$,
and ${\cal B}[\delta](z_i;k_i,\epsilon_i;p_I^\mu)$
is a holomorphic function ${\cal B}[\delta]$
on ${\cal M}_2$. The relative phases $\epsilon_\delta$
\footnote{The phases $\epsilon_\delta$ should not be confused with the
polarization tensors $\epsilon_i$ of the external states. Both notations
are standard, which is why they have been kept.}
can be determined by the requirement that
${\cal H} \equiv \sum_\delta\epsilon_\delta{\cal B}[\delta]$
transforms so that the expression ${\bf A}_{II}$ of (\ref{A1}) be modular invariant.
The summation over spin structures $\delta$ is the Gliozzi-Scherk-Olive projection,
and, physically, it is necessary to project out tachyonic states and insure space-time
supersymmetry.

\medskip

$\bullet$ Once the phases $\epsilon_\delta$ have been determined
by the $0$-point function,
we can consider the sums $\sum_\delta\epsilon_\delta{\cal B}[\delta](z_i;k_i,\epsilon_i;p_I^\mu)$
directly for $N\geq 1$. Then $\sum_\delta\epsilon_\delta{\cal B}[\delta](z_i;k_i,\epsilon_i;p_I^\mu)=0$
for $N\leq 3$, while for $N=4$, there exist forms ${\cal S}_j(z_i;k_i,\epsilon_i;p_I^\mu)$
which are scalars in $z_j$ and closed 1-forms in $z_i$ for $i\not=j$ so that
\bea
\label{calS}
\sum_\delta\epsilon_\delta{\cal B}[\delta](z_i;k_i,\epsilon_i;p_I^\mu)
-
\sum_{j=1}^4d\bar z_j\,\p_{\bar z_j}{\cal S}_j(z_i;k_i,\epsilon_i;p_I^\mu)
~ \in ~ \bigotimes_{i=1}^4\Lambda^{1,0}_{z_i}(\Sigma).
\eea
The form ${\cal H}(z_i;k_i,\epsilon_i;p_I^\mu)$ can now be obtained by
\bea
\label{calH1}
{\cal H}(z_i;k_i,\epsilon_i;p_I^\mu)
=
\sum_\delta\epsilon_\delta{\cal B}[\delta](z_i;k_i,\epsilon_i;p_I^\mu)
-
\sum_{j=1}^4d_j{\cal S}_j(z_i;k_i,\epsilon_i;p_I^\mu),
\eea
where $d_j$ is the de Rham exterior differential in each variable $z_j$.
The closedness of ${\cal B}[\delta](z_i;k_i,\epsilon_i;p_I^\mu)$ implies
that
${\cal H}(z_i;k_i,z_i;p_I^\mu)$ is automatically holomorphic in each $z_i$.

\medskip
We shall see below that the chiral forms ${\cal B}[\delta](z_i;k_i,\epsilon_i;p_I^\mu)$
arise from {\it superholomorphic} forms with respect to a supergeometry
$(g_{mn},\chi_m{}^\alpha)$. The above Dolbeault cohomology procedure
solves an old puzzle: there is no relation between superholomorphicity
and holomorphicity with respect to $g_{mn}$, but there is
a {\it deformed metric} $\hat g_{mn}$ with respect to which holomorphic forms
can be extracted from
superholomorphic forms modulo forms which are Dolbeault-exact in one
and de Rham-closed in the other insertion points.

\subsection{The forms ${\cal B}[\delta]$ in terms of Green's functions}

The amplitudes ${\cal B}[\delta](z_i;k_i,\epsilon_i;p_I^\mu)$ are to be determined by taking
the chiral contributions of functional integrals over all fluctuating worldsheets
and all insertion points $z_i$ for the emission of the $N$
massless bosons, and factoring out correctly the gauge symmetries to
arrive at well-defined, finite-dimensional integrals.

\medskip
The basic result is that,
by following the gauge-fixing procedure outlined in Section \S 3, the
${\cal B}[\delta](z_i;k_i,\epsilon_i;p_I^\mu)$ are found to be \cite{V}
\bea
\label{BdBc}
{\cal B}[\delta]={\cal B}[\delta]^{(d)}+{\cal B}[\delta]^{(c)}.
\eea
Here the ``connected" and ``disconnected" components ${\cal B}[\delta]^{(d)}$
and ${\cal B}[\delta]^{(c)}$ are given in terms of two basic measures
$d\mu_2[\delta]$ and $d\mu_0[\delta]$ on the moduli space ${\cal M}_2$
and Wick contractions of vertex operators ${\cal V}^{(0)}$,
${\cal V}^{(1)}$, and ${\cal V}^{(2)}$. The vertex operators
${\cal V}^{(0)}$, ${\cal V}^{(1)}$, ${\cal V}^{(2)}$ are defined by
\bea
\label{vertex}
{\cal V}^{(0)}(z)&=& \epsilon^\mu dz\,(\p_zx_+^\mu-ik^\nu\psi_+^\mu\psi_+^\nu)(z)
{\rm exp}(ik\cdot x_+(z))
\nonumber\\
{\cal V}^{(1)}(z)&=&-{1\over 2}\epsilon^\mu d\bar z\,\chiz \psi_+^\mu(z)
{\rm exp}(ik\cdot x_+(z))
\nonumber\\
{\cal V}^{(2)}(z)&=&
-\epsilon^\mu \hat \mu_{\bar z}{}^z d\bar z\,(\p_zx_+^\mu
-ik^\nu\psi_+^\mu\psi_+^\nu)(z)
{\rm exp}(ik\cdot x_+(z))
\eea
where $x_+^\mu$ is an effective chiral scalar field with propagator
$\<x_+^\mu (z)x_+^\nu (w)\>=-\delta^{\mu\nu}\ln E(z,w)$, $E(z,w)$ being
the prime form on the
Riemann surface $\Sigma$.
The spin structure $\delta$ determines a square root $\Lambda^{{1\over 2},0}[\delta](\Sigma)$
of the canonical bundle of $\Sigma$. The gravitino field $\chi(z)=(\chi_m{}^\alpha)$
is a section of $\Lambda^{0,1}(\Sigma)\otimes \Lambda^{-{1\over 2},0}[\delta](\Sigma)$.
It is given by $\chi(z)=\sum_{\alpha=1}^2\zeta^\alpha\chi_\alpha(z)$
where $\chi_\alpha(z)$ are two fixed, generic, but otherwise arbitrary
sections of $\Lambda^{0,1}(\Sigma)\otimes \Lambda^{-{1\over 2},0}[\delta](\Sigma)$,
$\zeta^\alpha$ are two
anti-commuting parameters (corresponding to the odd supermoduli of
$s{\cal M}_2$ to be discussed in Section \S 3), and $\hat \mu(z)=
(\hat \mu_{\bar z}{}^z)\in \Lambda^{-1,1}(\Sigma)$ is a Beltrami differential
defined modulo $\bar\partial\,T^{1,0}(\Sigma)$ by
the condition
\bea
\label{chiandmu}
{1\over 8\pi}\int_\Sigma\int_\Sigma d^2zd^2w\,
\omega_I(z)\chi(z) S_\delta(z,w)
\chi(w) \omega_J(w)
=
\int_\Sigma \omega_I(z)\omega_J(z)\, \hat \mu(z).
\eea
where $S_\delta(z,w)$ is the Szeg\"o kernel.
The measures $d\mu_2[\delta]$ and $d\mu_0[\delta]$
on ${\cal M}_2$ are defined by
\bea
\label{measures}
d\mu_0[\delta](\Omega)
&=&
{\cal Z}[\delta]\
\prod_{I\leq J} d\Omega_{IJ}
\nonumber\\
d\mu_2[\delta](\Omega)
&=&
{\cal Z}[\delta]\
\sum_{j=1}^6
{\cal X}_j\
\prod_{I\leq J} d\Omega_{IJ}
\eea
with the following expressions for ${\cal Z}[\delta]$ and ${\cal X}_j$, $1\leq j\leq 6$:
\bea
\label{calZgeneral}
{\cal Z}[\delta]=
{\<\prod_a b(p_a)\prod_{\alpha}\delta(\beta(q_\alpha))\>
\over {\rm det}\,(\omega_I\omega_J(p_a))\cdot
\<\chi_\alpha|\psi_\beta^*\>},
\eea
where $p_a, q_\alpha$ are two sets of respectively 3 and 2 arbitrary generic
points, and $\psi_\beta^*$ are the holomorphic forms of weight
$3/2$ normalized at the points $q_\alpha$ by $\psi_\beta^*(q_\alpha)=\delta_{\alpha\beta}$.
The fields $b(z)=b_{zz},\beta(z)=\beta_{z+}$ and their partners
$c(z)=c^z$, $\gamma(z)=\gamma^+$ are the so-called superghost
fields,
with propagators
\bea
\<b(z)c(w)\>=G_2(z,w),
\qquad
\<\beta(z)\gamma(w)\>=-G_{3/2}(z,w)
\eea
where $G_n(z,w)$ are the Green's functions on tensors of weight $n$.
Next, let $S(z)=S_{z+}$ and $T(z)=T_{zz}$ be the supercurrent and
the stress tensor defined by
\bea
S(z)&=&-{1\over 2}\psi_+^\mu\p_zx_+^\mu
+
{1\over 2}b\gamma-{3\over 2}\beta\p_zc-(\p_z\beta)c
\nonumber\\
T(z)&=&
-{1\over 2}\p_zx^\mu\p_zx^\mu
+
{1\over 2}\psi_+^\mu\p_z\psi_+^\mu
+c\p_zb+2(\p_zc)b
-{1\over 2}\gamma\p_z\beta
-{3\over 2}(\p_z\gamma)\beta.
\eea
Then the expressions ${\cal X}_j$, $1\leq j\leq 6$, are given by
\bea
\label{X}
{\cal X}_1&=&
-{1\over 8\pi^2}
\int d^2z \chiz\int d^2w \chiw\,\<S(z)S(w)\>
\nonumber\\
{\cal X}_2+{\cal X}_3
&=&+{1\over 16\pi^2}
\int d^2z\int d^2w\, \chiz\chiw T^{IJ}\omega_I(z)S_\delta(z,w)\omega_J(w)
\nonumber\\
{\cal X}_4
&=&
+{1\over 16\pi^2}\int d^2w\ \p_{p_a}\p_w\ln E(p_a,w)\chiw \int d^2u
S_\delta(w,u)\chi_{\bar u}{}^+\varpi_a^*(u)
\nonumber\\
{\cal X}_5
&=&
+{1\over 16\pi^2}\int d^2u\int d^2v \ S_\delta(p_a,u)
\chi_{\bar u}{}^+\p_{p_a}S_\delta(p_a,v)\chi_{\bar v}{}^+\varpi_a(u,v)
\nonumber\\
{\cal X}_6&=&
{1\over 16\pi^2}\int d^2z \chi_\alpha^*(z)
\int d^2w \,G_{3/2}(z,w)\chiw\int d^2v \chiv \Lambda_\alpha(w,v)
\eea
where $\Lambda_\alpha(w,v)=2G_2(w,v)\p_v\psi_\alpha^*
+3\p_vG_2(w,v)\psi_\alpha^*(v)$, the sections $\chi\beta^*(z)$
are the linear combinations of the sections $\chi_\alpha(z)$
normalized by $\<\chi_\beta^*|\psi_\alpha^*\>
=\delta_{\alpha\beta}$, and $T^{IJ}$ are
the coefficients of the holomorphic quadratic differential defined by
\bea
\label{TIJ}
T^{IJ}\omega_I\omega_J(w)
&=& {\<T(w)\prod_{a=1}^3b(p_a)\prod_{\alpha=1}^2\delta(\beta(q_\alpha))>
\over \<\prod_{a=1}^3b(p_a)\prod_{\alpha=1}^2\delta(\beta(q_\alpha))\>}
-2\sum_{a=1}^3\p_{p_a}\p_w\ln E(p_a,w)\varpi_a^*(w)
\nonumber\\
&&
\quad
+
\int d^2z\,\chi_\alpha^*(z)
\big(-{3\over 2}\p_wG_{3/2}(z,w)\psi_\alpha^*(w)
-
{1\over 2}G_{3/2}(z,w)\p\psi_\alpha^*(w)
\nonumber\\
&&
\quad\qquad
+G_2(w,z)\p_z\psi_\alpha^*(z)
+{3\over 2}\p_zG_2(w,z)\psi_\alpha^*(z)\big),
\eea
and $\varpi_a^*$ and $\varpi_a$ are holomorphic forms in $u$ and $v$
defined by $\varpi_a^*(u)=\varpi_a(u,p_a)$ and
\bea
\varpi_a(u,v)&=& {{\rm det}\{\omega_I\omega_J(p_b[u,v;a])\}
\over {\rm det}\{\omega_I\omega_J(p_b)\}}
\nonumber\\
\omega_I\omega_J(p_b[u,v;a])
&=& \cases{\omega_I\omega_J(p_b) & if $b\not=a$\cr
{1\over 2}(\omega_I(u)\omega_J(v)+\omega_I(v)\omega_J(u))
&if $b=a$\cr}
\eea
In (\ref{TIJ}), all the apparent poles cancel, which is why $T^{IJ}$
is well-defined. In the expressions for $\varpi_a^*$ and $\varpi_a$,
the indices $IJ$ and $a$ are both $3$-dimensional, and hence it makes
sense to take the $3\times 3$ determinants indicated.

\medskip

We can now give the expressions for ${\cal B}[\delta]^{(c)}$ and
${\cal B}[\delta]^{(1)}$,
\bea
{\cal B}[\delta]^{(d)}&=&
d\mu_2[\delta]\, \big\<Q(p_I)\prod_{i=1}^N{\cal V}_i^{(0)}(z_i;k_i)\big\>
\nonumber\\
{\cal B}[\delta]^{(c)}&=&
d\mu_0[\delta]\int \prod_{\alpha=1}^2d\zeta^\alpha\sum_{j=1}^5{\cal Y}_j,
\eea
where $Q(p_I)={\rm exp}(ip_I^\nu\oint_{B_I}dz\,\p_zx_+^\nu(z))$, and
\bea
\label{Y}
{\cal Y}_1&=&{1\over 8\pi^2}\<Q(p_I)\int \chi S\int \chi S\prod_{j=1}^N{\cal V}_j^{(0)}\>_{(c)}
\nonumber\\
{\cal Y}_2&=&
{1\over 2\pi}
\<Q(p_I)\int \hat \mu T\prod_{j=1}^N{\cal V}_j^{(0)}\>
\nonumber\\
{\cal Y}_3&=&{1\over 2\pi}
\sum_{i=1}^N\<Q(p_I)\int \chi S{\cal V}_i^{(1)}\prod_{j\not=i}{\cal V}_j^{(0)}\>
\nonumber\\
{\cal Y}_4&=&{1\over 2}\<Q(p_I){\cal V}_i^{(1)}{\cal V}_j^{(1)}
\prod_{l\not= i,j}{\cal V}_l^{(0)}\>
\nonumber\\
{\cal Y}_5&=&
\sum_{i=1}^N\<Q(p_I){\cal V}_i^{(2)}\prod_{j\not=i}{\cal V}_j^{(0)}\>
\eea

The preceding formulas give a complete and systematic way of
obtaining the scattering amplitude for $N$ massless bosons to two-loop
order. Their interpretation is roughly as follows.
The choice $\chi(z)=\sum_{\alpha=1}^2\zeta^\alpha\chi_{\alpha}(z)$
is a choice of gauge slice.
The fundamental guiding principle of our gauge-fixing method is to
project the supergeometry $(g_{mn},\chi_m{}^\alpha)$ on a super period
matrix invariant under supersymmetry, rather than on the metric $g_{mn}$.
Since the functional integrals are originally defined in terms of
the metric $g_{mn}$, this requires a deformation of complex structures
implemented through the Beltrami differential $\hat \mu(z)$. The terms ${\cal X}_j$,
$2\leq j\leq 5$, incorporate both local and global effects of this deformation
of complex structures. In general, the emission of a string state is implemented
by insertion of a vertex, in this case,
the vertex ${\cal V}^{(0)}$ which is the naive vertex for graviton emission. However,
due to the gauge-fixing procedure and the deformation of
complex structures, the naive vertex must be corrected by the
vertices ${\cal V}^{(1)}$ and ${\cal V}^{(2)}$. This produces the terms
${\cal Y}_j$, $2\leq j\leq 5$.
Note that ${\cal V}^{(0)}$ is a $(1,0)$-form,
but ${\cal V}^{(1)}$, ${\cal V}^{(2)}$ are $(0,1)$-forms.
The period matrix $\Omega_{IJ}$ of the previous formulas is actually
the period matrix $\hat\Omega_{IJ}$ of the metric $\hat g_{mn}$, but {\it after}
the deformation of complex structures, we drop the ``hat" notation for simplicity.

\subsection{Gauge-slice independence of the measure $d\mu_2[\delta](\Omega)$}

Next, the amplitudes ${\bf A}_{II}$ have to be shown to be
independent of all the choices of $q_\alpha$, $p_a$,
$\chi_\alpha(z)$, $\mu(z)$ entering the amplitudes ${\cal
B}[\delta](z_i;k_i,\epsilon_i;k_I^\mu)$. This is important because
it had not been satisfied by earlier Ans\"atze, and there had been
concern that superstring scattering amplitudes could be ambiguous.
It also paves the way for the evaluation of ${\cal
B}[\delta](z_i;k_i,\epsilon_i;p_I^\mu)$ in terms of
$\tet$-functions.

\medskip

We begin with the measure $d\mu_2[\delta](\Omega)$ \cite{II,III}. The gauge slice independence
of $d\mu_2[\delta](\Omega)$ is established by showing that its variational
derivative with respect to
any of the above choices vanishes identically on the moduli space
${\cal M}_2$ \cite{II}. The following special case is of considerable practical value,
and produces relatively simpler expressions which can independently be shown
to be independent of all remaining choices.
Choose $\chi_\alpha(z)$
to be a Dirac measure at a point $x_\alpha$ and let $x_\alpha\to q_\alpha$. All dependence on
$\mu(z)$ cancels
out completely, and the resulting expression for $d\mu_2[\delta]$ becomes
\bea
\label{mu2dirac}
d\mu_2[\delta]
=
{\cal Z}[\delta]\
\sum_{j=1}^6{\cal X}_j,
\eea
with
\bea
{\cal Z}[\delta]={\<\prod_{a=1}^3b(p_a)\prod_{\alpha=1}^2\delta(\beta(q_\alpha))\>
\over
{\rm det}\,\omega_I\omega_J(p_a)}
\eea
and the terms ${\cal X}_i$ given by,
\bea
\label{calX}
{\cal X}_1
+{\cal X}_6
&=& {\zeta^1\zeta^2\over 16\pi^2}[-10 S_\delta(q_1,q_2)\p_{q_1}\p_{q_2}\ln E(q_1,q_2)
\nonumber\\
&&
\qquad\quad
-\p_{q_1}G_2(q_1,q_2)\p\psi_1^*(q_2)+\p_{q_2}G_2(q_2,q_1)\p\psi_2^*(q_1)
\nonumber\\
&&
\qquad\quad
+2G_2(q_1,q_2)\p\psi_1^*(q_2)f_{3/2}^{(1)}(q_2)
-
2G_2(q_2,q_1)\p\psi_2^*(q_1)f_{3/2}^{(2)}(q_1)]
\nonumber\\
{\cal X}_2
&=&
 {\zeta^1\zeta^2\over 16\pi^2}
\omega_I(q_1)\omega_J(q_2)S_\delta(q_1,q_2)
[\p_I\p_J\ln {\tet[\delta](0)^5\over\tet[\delta](D_\beta)}
+
\p_I\p_J\ln \tet(D_b)]
\nonumber\\
{\cal X}_3&=&
{\zeta^1\zeta^2\over 8\pi^2}
S_\delta(q_1,q_2)\sum_a\varpi_a(q_1,q_2)[B_2(p_a)+B_{3/2}(p_a)]
\nonumber\\
{\cal X}_4&=&
{\zeta^1\zeta^2\over 8\pi^2}
S_\delta(q_1,q_2)\sum_a[\p_{p_a}\p_{q_1}\ln E(p_a,q_1)\varpi_a^*(q_2)
+
\p_{p_a}\p_{q_2}\ln E(p_a,q_2)\varpi_a^*(q_1)]
\nonumber\\
{\cal X}_5&=&
{\zeta^1\zeta^2\over 16\pi^2}
\sum_a[S_\delta(p_a,q_1)\p_{p_a}S_\delta(p_a,q_2)
-
S_\delta(p_a,q_2)\p_{p_a}S_\delta(p_a,q_1)]\varpi_a(q_1,q_2)
\eea
Here $D_b=p_1+p_2+p_3-3\Delta$, $D_\beta=q_1+q_2-2\Delta$,
and the expressions
$f_n(w)$, $f_{3/2}^{(1)}(x)$, $f_{3/2}^{(2)}(x)$,
$B_2(w)$ and $B_{3/2}(w)$ are given by
\bea
f_n(w)
&=& \omega_I(w)\p_I\ln \tet[\delta](D_n)
+\p_w\ln (\sigma(w)^{2n-1}\prod_{i=1}^{2n-1}E(w,z_i))
\nonumber\\
f_{3/2}^{(1)}(x)
&=&
\omega_I(q_1)
\p_I\ln \tet[\delta](x+q_2-2\Delta)
+
\p_{q_1}\ln (E(q_1,q_2)E(q_1,x)\sigma(q_1)^2)
\nonumber\\
f_{3/2}^{(2)}(x)
&=&
\omega_I(q_2)
\p_I\ln \tet[\delta](x+q_1-2\Delta)
+
\p_{q_2}\ln (E(q_2,q_1)E(q_2,x)\sigma(q_2)^2)
\nonumber\\
B_2(w)
&=& -2 T_1(w)
+{1\over 2}f_2(w)^2
-{3\over 2}\p_wf_2(w)
-
2\sum_a\p_{p_a}\p_w\ln E(p_a,w)\varpi_a^*(w)
\nonumber\\
B_{3/2}(w)
&=& 12\,T_1(w)-{1\over 2}f_{3/2}(w)^2
+\p f_{3/2}(w)
\eea
with $\Delta$ the vector of Riemann constants, $\sigma(z)$ the basic function
with monodromy introduced in \cite{vv87-2, fay, faltings}, and $E(z,w)=(z-w)
+(z-w)^2T_1(w)+O((z-w)^3)$ defines the chiral scalar bosonic stress tensor $-T_1(w)$.

\medskip
Compared with the earlier expression (\ref{X}) for ${\cal X}_j$
and for $d\mu_2[\delta]$, all field theoretic
correlation functions have been worked out, and the new expression only
involves complex function theory on the Riemann surface $\Sigma$.
It can be checked directly to be independent of the choice
points $p_a$, $q_\alpha$ \cite{III}.

\medskip
The measure $d\mu_2[\delta]$ suffices to determine the $N=0$ amplitude, which
is also the space-time cosmological constant. In fact, in this case, there is
no vertex operator, and the internal momenta $p_I^\mu$ can be integrated out
to give
\bea
{\bf A}_{II}\bigg|_{N=0}=\int_{{\cal M}_2}({\rm det}\,\Im\,\Omega)^{-5}
\
\sum_\delta
\epsilon_\delta d\mu_2[\delta](\Omega)
\wedge
\overline{
\sum_\delta
\epsilon_\delta d\mu_2[\delta](\Omega)},
\eea
with the phases $\epsilon_\delta$ yet to be determined by modular invariance.

\subsection{Gauge-slice independence of the $N$-point function}

We consider next the slice-independence of the $N$-point function
\cite{V}.
Since the correlator $\<Q(p_I)\prod_{j=1}^N{\cal V}^{(0)}(z_j)\>$
is manifestly independent of any choice of gauge-slice
and since $d\mu_2[\delta]$ has been shown to be slice-independent,
the term ${\cal B}[\delta]^{(d)}$ is slice-independent.

\medskip

The term ${\cal B}[\delta]^{(c)}$ is not invariant under changes of
gauge slices, but it transforms by \cite{V}
\bea
\label{btransform}
{\cal B}[\delta]^{(c)}(z_i;k_i,\epsilon_i;p_I)
\to
{\cal B}[\delta]^{(c)}(z_i;k_i,\epsilon_i;p_I)
+
\sum_{i=1}^N d_i{\cal R}_i[\delta](z_i;k_i,\epsilon_i;p_I),
\eea
where the forms ${\cal R}_i[\delta](z_i;k_i,\epsilon_i;p_I^\mu)$ are scalars in $z_i$,
de Rham closed forms
in $z_j$ for $j\not=i$, and have the same monodromy as ${\cal B}[\delta]$.
Since the forms ${\cal B}[\delta]$ are closed in each $z_i$, and since by
analytic continuation \cite{dp92}, the singularities at coincident insertion points $z_i=z_j$
are harmless, it follows from a Riemann bilinear relations argument that
the terms ${\cal R}_i[\delta]$ do not contribute to the integrated
amplitudes ${\bf A}_{II}(k_i,\epsilon_i)$.
Thus the $N$-point functions ${\bf A}_{II}(k_i,\epsilon_i)$ are gauge slice-independent.

\subsection{The measure $d\mu_2[\delta]$ and the
modular covariant form $\Xi_6[\delta]$}

Once the gauge slice independence has been established, the chiral amplitudes
${\cal B}[\delta]$ can be evaluated explicitly by making convenient choices
for the points $p_a$, $q_\alpha$.

\medskip
The first fundamental term is $d\mu_2[\delta](\Omega)$, which is the chiral string
measure, and which will determine the phases $\epsilon_\delta$. We find \cite{I,IV}
\bea
d\mu_2[\delta](\Omega)
=
{1\over 16\pi^6}{\Xi_6[\delta](\Omega)\tet[\delta](\Omega)^4
\over
\Psi_{10}(\Omega)}\prod_{I\leq J}d\Omega_{IJ}
\eea
The form $\Psi_{10}(\Omega)$ is the familiar modular form of weight 10
defined by
\bea
\Psi_{10}(\Omega)=\prod_{\delta\ even}\tet[\delta](\Omega)^2.
\eea
The key new form is $\Xi_6[\delta](\Omega)$, whose construction depends on some
particular properties of even spin structures in genus $h=2$.
Recall that, in genus $h=2$, there are 10 even spin structures $\delta$
and 6 odd spin structures $\nu$, denoted by $\nu_1,\cdots,\nu_6$. Any even spin structure $\delta$
can be decomposed as a sum of 3 odd spin structures. If we write
$\delta$ accordingly as $\delta=\nu_1+\nu_2+\nu_3$, then $\Xi_6[\delta](\Omega)$
is given by
\bea
\label{Xi6}
\Xi_6[\delta]
=
\sum_{1\leq i<j\leq 3}\<\nu_i|\nu_j\>\prod_{k=4,5,6}
\tet[\nu_i+\nu_j+\nu_k]^4(\Omega).
\eea
A very important property of $\Xi_6[\delta](\Omega)$ is its transformation law under
$Sp(4,{\bf Z})$,
which is not quite that a modular form, but rather
\bea
\Xi_6[\tilde\delta](\tilde\Omega)
=\epsilon^4{\rm det}\,(C\Omega+D)^2\Xi_6[\delta](\Omega),
\qquad \pmatrix{A & B\cr C&D\cr}\in Sp(4,{\bf Z}),
\eea
where $\tilde\Omega=(A\Omega+B)(C\Omega+D)^{-1}$,
$\tilde\delta$ is the corresponding transform of the spin structure
$\delta$, and $\epsilon$ is exactly the same 8th-root of unity which occurs
in the transformation law for $\tet$-constants,
$\tet[\tilde\delta](\tilde\Omega)
=\epsilon^4{\rm det}(C\Omega+D)^6\tet[\delta](\Omega)$.
There would have been no such factors $\epsilon^4$ in the transformation law for
modular forms. This shows that there is a unique choice of relative phases
$\epsilon_\delta=+1$
between the various even spin structures for the GSO projection, given by
$\sum_\delta d\mu[\delta](\Omega)$. By examining degenerations of
the surface $\Sigma$, it is then not difficult to show that
\bea
\label{identity1}
\sum_\delta\Xi_6[\delta](\Omega)\tet[\delta](\Omega)^4=0,
\eea
and hence $\sum_\delta d\mu_2[\delta](\Omega)=0$.
Physically, this means that the cosmological
constant vanishes in superstring theory, which is a consequence of
space-time supersymmetry. Mathematically, for genus $h=1$, the vanishing
of the cosmological constant was known to follow from
the Jacobi identity for $\tet$-constants, and thus from the Riemann
identities. In genus 2 however,
the identity (\ref{identity1}) does not follow
from the Riemann identities alone. Rather, it is equivalent to
the fact that an $Sp(4,{\bf Z})$ modular form of weight 8 must be proportional
to the square of the unique $Sp(4,{\bf Z})$ modular form of weight 4.

\subsection{Explicit formula for the holomorphic form ${\cal H}$}

Once the relative phases $\epsilon_\delta=1$ have been determined,
we can evaluate directly the Gliozzi-Scherk-Olive sum
$\sum_\delta \epsilon_\delta {\cal B}[\delta]=\sum_\delta{\cal B}[\delta]$
instead of evaluating each ${\cal B}[\delta]$ separately. Using now the
unitary gauge with $q_\alpha$ the divisor
of a holomorphic one form $\varpi(z)$, we find \cite{VI}
\bea
\sum_\delta {\cal B}[\delta](z_i;k_i,\epsilon_i;p_I^\mu)
=0,
\qquad 0\leq N\leq 3,
\eea
while for $N=4$, we find
\bea
\sum_\delta {\cal B}[\delta](z_i;k_i,\epsilon_i;p_I^\mu)
=
{\cal H}(z_i;k_i,\epsilon_i;p_I^\mu)
+\sum d_i\bigg(\Lambda(z_i)\<Q(p_I)
\prod_{j=1}^4e^{ik_jx_+(z_j)}\>\prod_{j\not=i}\varpi(z_j)\bigg)
\eea
where $\Lambda(z)$ is a certain single-valued smooth scalar function, and
the holomorphic form ${\cal H}(z_i;k_i,\epsilon_i;p_I^\mu)$ is given by
\bea
{\cal H}(z_i;k_i,\epsilon_i;p_I^\mu)={1\over 64\pi^2}K{\cal Y}_S
{\rm exp}(i\pi p_I^\mu\Omega_{IJ}p_J^\mu
+2\pi i\sum_{j=1}^4p_I^\mu k_j^\mu\int^{z_j}\omega_I)
\prod_{i<j}E(z_i,z_j)^{k_i\cdot k_j}
\eea
where the factor ${\cal Y}_S$ is defined to be
\bea
3\,{\cal Y}_S&=&+(k_1-k_2)\cdot(k_3-k_4)\,\Delta(z_1,z_2)\Delta(z_3,z_4)\nonumber\\
&&
+(k_1-k_3)\cdot(k_2-k_4)\,\Delta(z_1,z_3)\Delta(z_2,z_4)\nonumber\\
&&
+(k_1-k_4)\cdot(k_2-k_3)\,\Delta(z_1,z_4)\Delta(z_2,z_3)
\eea
and $\Delta(z,w)=\omega_1(z)\omega_2(w)-\omega_1(w)\omega_2(z)$ is the basic
anti-symmetric biholomorphic form.
The kinematic factor $K=K(1,2,3,4)$ is the same one as in tree-level and one-loop amplitudes.
Explicitly, in terms of the gauge-invariant field strengths $f_i^{\mu\nu}
=\epsilon_i^\mu k_i^\nu-\epsilon_i^\nu k_i^\mu$, it can be written as
\bea
K(1,2,3,4)&=& (f_1f_2)(f_3f_4)+(f_1f_3)(f_2f_4)+(f_1f_4)(f_2f_3)
\nonumber\\
&&-4(f_1f_2f_3f_4)-4(f_1f_3f_2f_4)-4(f_1f_2f_4f_3),
\eea
with $(f_if_j)=f_i^{\mu\nu}f_j^{\nu\mu}$,
$(f_if_jf_kf_l)=f_i^{\mu\nu}f_j^{\nu\rho}f_k^{\rho\sigma}f_l^{\sigma\mu}$.

\subsection{The $4$-point function}

Using equation (\ref{A1}), the 4-point function ${\bf A}_{II}$ follows readily from the
exact formula for ${\cal H}(z_i;k_i,\epsilon_i;p_I^\mu)$ which we just obtained.
The integral over the
internal momenta $p_I^\mu$ completes the factors $E(z_i,z_j)$ into Green's
functions, and we obtain \cite{VI}
\bea
\label{AII}
{\bf A}_{II}(k_i,\epsilon_i)
=
{K\bar K\over 2^{12}\pi^4}
\int_{{\cal M}_2\times\Sigma^4}
{|\prod_{I\leq J}d\Omega_{IJ}|^2
\over
({\rm det\,Im}\,\Omega)^{5}}
|{\cal Y}_S|^2
{\rm exp}(-\sum_{i<j}k_i\cdot k_j G(z_i,z_j))
\eea
where $G(z,w)$ is the conformally invariant Green's function
\bea
G(z,w)=-\ln |E(z,w)|^2+2\pi({\rm Im}\,\Omega)_{IJ}^{-1}
({\rm Im}\int_z^w\omega_I)({\rm Im}\int_z^w\omega_J).
\eea
An expression in the hyperelliptic representation equivalent to (\ref{AII})
was partly guessed in \cite{zwz}, starting also
from the measures $d\mu_2[\delta]$ and
$d\mu_0[\delta]$ given in \cite{I,II,III,IV}.
The derivation in \cite{zwz}
is however not gauge slice independent,
because the corrections ${\cal V}^{(1)}$
and ${\cal V}^{(2)}$ to the vertex operators were not taken into
account.

The 4-point functions for the heterotic string are obtained by
replacing in (\ref{A1}), at common loop momenta $p_I^\mu$, the holomorphic
factors by the holomorphic blocks of the 10-dimensional bosonic
string coupled with 32 worldsheet chiral fermions.
They are of the form
\bea
\label{AHET}
{\bf A}_{HET}
&=&
{K\bar K\over 2^{12}\pi^4}
\int_{{\cal M}_2\times\Sigma^4}
{|\prod_{I\leq J}d\Omega_{IJ}|^2
\over \pi^{12}\Psi_{10}(\Omega)
({\rm det\,Im}\,\Omega)^{5}}
{\cal W}(z_1,z_2,z_3,z_4)\overline{{\cal Y}_S(z_1,z_2,z_3,z_4)}
\nonumber\\
&&
\qquad\qquad\qquad
\times{\rm exp}\bigg(-\sum_{i<j}k_i\cdot k_j \,G(z_i,z_j)\bigg)
\eea
where the holomorphic block ${\cal W}(z_1,z_2,z_3,z_4)$ depends on whether
on the external states and can be written down explicitly. For example,
for the relatively more complicated scattering of two gravitons and two gauge bosons,
we have respectively ${\cal W}={\cal W}_{(R^2F^2)}$ and ${\cal W}={\cal W}_{(R^4)}$, with
\bea
\label{calW}
{\cal W}_{(R^2F^2)}
&=& {\cal W}_{(F^2)}(z_1,z_2)\{\epsilon_1^\mu\epsilon_2^\mu
\p_{z_3}\p_{z_4}G(z_3,z_4)-\sum_{ij}\epsilon_3^\mu k_i^\mu\epsilon_4^\nu k_j^\nu
\p_{z_3}G(z_3,z_i)\p_{z_4}G(z_4,z_j)\}
\nonumber\\
{\cal W}_ {(R^4)}
&=&{\<\prod_{j=1}^4\epsilon_j^\mu\p x^\mu(z_j)e^{ik_j\cdot x(z_j)}\>
\over
\<\prod_{j=1}^4 e^{ik_j\cdot x(z_j)}\>}
\eea
where $x(z,\bar z)$ is a non-chiral scalar field with propagator $G(z,w)$, and
\bea
{\cal W}_{(F^2)}(z_1,z_2)&=&{1\over 2}tr(T^{a_1}T^{a_2})\sum_\kappa \tet[\kappa]^8S_\kappa(z_1,z_2)^2
\nonumber\\
{\cal W}_{(F^2)}(z_1,z_2)&=&{1\over 2}tr (T^{a_1}T^{a_2})\sum_\kappa\tet[\kappa]^8
\sum_\rho \tet[\rho]^4S_\rho(z_1,z_2)^2,
\eea
depending on whether the heterotic theory is the $Spin(32)/{\bf Z}_2$ or
the $E_8\times E_8$ theory.

\subsection{Non-renormalization theorems}

The low-energy effective action of superstring theories provides corrections
to the Einstein action involving higher order curvature terms as well
as couplings to additional fields such as gauge bosons
\cite{Gross:1986iv}. The amplitudes ${\bf A}_{II}$,
${\bf A}_{HET}$ we just obtained allow us to determine readily the two-loop corrections to
terms such as $\R^4$ in the type II superstring, and $F^4$, $F^2F^2$,
$\R^2F^2$, $\R^4$
in the heterotic strings \cite{VI}. Here $\R^4 = t_8 t_8 R^4$, and $R$ is the
space-time Riemann curvature tensor, and
$F$ is the curvature of the gauge bosons. In determining the low-energy corrections,
we have to let $k_i\to 0$, but only after the amplitude has been expressed in terms
of the field strengths
$f_i^{\mu\nu}=\epsilon_i^\mu k_i^\nu-\epsilon_i^\nu k_i^\mu$.
A strong motivation for determining these corrections are the conjectured
dualities between the $Spin(32)/{\bf Z}_2$ heterotic theory and the type I superstring,
as well as the S-duality of the type IIB superstring (see the next section).

\medskip
For the type II superstring, it is manifest from the explicit form of
${\bf A}_{II}(k_i,\epsilon_i)$ that
the two-loop contribution to $\R^4$ vanishes. The heterotic strings are more subtle,
because the contributions of the bosonic left sector necessarily have poles in
the Mandelstam variables $s_{ij}=-2k_i\cdot k_j$. Nevertheless, we find that
terms such as $s_{ij}{\cal W}_{(\R^2)}$ and $s_{ij}s_{lm}{\cal W}_{(\R^4)}$
can be expressed in expressions such as
\bea
\p_{z_1}\p_{z_2}G(z_1,z_2)\,{\rm exp}(-\sum_{i<j}k_i\cdot k_j G(z_i,z_j)),
\qquad
\sum_{i<j}C_{ij}^{\mu\nu}\p_{z_1}G(z_1,z_i)\p_{z_2}G(z_2,z_j)
\eea
whose integrals against holomorphic differentials tend to 0 as $k_i\cdot k_j\to 0$.
This turns out to suffice to establish the desired non-renormalization
theorem, by which the terms $\R^2F^2$ and $\R^4$ in the heterotic string
do not receive corrections to two-loop order \cite{VI}.

\subsection{S-duality for the type IIB superstring}

Here we discuss joint work with M. Gutperle on a partial check of the famous
$SL(2,{\bf Z})$ dualities for the type IIB superstring conjectured by M. Green, M. Gutperle, P. Vanhove, H.G. Kwon, and others
(see \cite{Green:1997as,conjecture,Green:2005ba},
and references in \cite{dgp}).
$S$-duality provides powerful
constraints on the form of the low-energy effective actions.
In particular, it was conjectured in \cite{conjecture}
that the $D^4\R^4$ terms in the type IIB effective action are of the form
\bea
S_{D^4R^4}=C_{D^4\R^4}\int d^{10}x\sqrt{-G}\,D^4{\cal R}^4 e^{{1\over 2}\phi}
2\zeta(5)E_{5/2}(\tau,\bar\tau)
\eea
where $\tau=\chi+ie^{-\phi}$ is the axion/dilaton field,
 $\zeta(s)$ is the Riemann zeta function,
and $E_{5/2}(\tau,\bar\tau)$ is the non-holomorphic
Eisenstein series of weight $s=5/2$,
\bea
2\zeta(s) E_s(\tau,\bar\tau)=\sum_{(m,n)\not=(0,0)}{\tau_2^s
\over|m+n\tau|^{2s}}.
\eea
Expanding $2\zeta(5)E_{5/2}(\tau,\bar\tau)$ in $\tau$, this conjecture predicts in
particular the precise value of the contribution to the $D^4\R^4$ of the two-loop
perturbative amplitude.

\medskip
This prediction can be compared with that of the formula (\ref{AII}), which gives the two-loop
amplitude up to an overall constant due to bosonization
formulas. The precise value of this
constant can be determined using factorization. We find that it matches exactly
that predicted from Eisenstein series, and thus the perturbative two-loop
amplitude provides a partial confirmation
of the conjectured S-duality \cite{dgp}.

\subsection{Orbifolds and Kachru-Kumar-Silverstein models}

So far, we have considered only superstrings evolving in flat Minkowski
space-time. However, the preceding gauge-fixing procedure adapts
readily to
other space-times, simply by replacing the correlation
functions of the fields $x_+^\mu$, $\psi_+^\mu$ by
those of the corresponding conformal field theory \cite{I}. Here we discuss joint
work with K. Aoki on the cosmological constant of some orbifold models
proposed by S. Kachru, S. Kumar, and E. Silverstein \cite{kks}.
These KKS models are of particular interest since their supersymmetry is broken,
yet their cosmological constant vanishes to one-loop. There was initially
some hope that the cosmological constant would still vanish to two loops,
but we can now show, using the new gauge-fixing method, that this is not
the case \cite{adp}.

\medskip
The KKS models are constructed with an orbifold group $G$ generated by
two elements $f=((r_L,s_R)^{1-4}, (1,s_R^2)^5,(s_L,s_R)^6;(-)^{F_R})$,
$g=(s_L,s_R)^{1-4}, (s_L,s_R)^5, (s_L^2,1)^6;(-)^{F_L})$ acting on a square torus
with self-dual radius. Here $s_L,s_R,r_L,r_R$ are chiral and reflections acting
on the left and right sectors, and the superscripts denote the dimension on which
the operator acts. The orbifold action creates sectors for the theory, indexed by
two twists $\epsilon$, $\alpha$, and the chiral string measure
$d\mu_2[\delta]$ is replaced now in each $(\epsilon,\alpha)$ sector by the
following measure,
\bea
d\mu_C[\delta;\epsilon,\alpha](p_L)=
{e^{i\pi \tau_\epsilon p_L^2}\over 16\pi^6\Psi_{10}}
{\tet[\delta_j^+]^2\tet[\delta_j^-]^2
\over \tet_j^4(0,\tau_\epsilon)}
\sum_\delta
\<\alpha|\delta\>\Xi_6[\delta]\tet[\delta]^2
\tet[\delta+\epsilon]^2.
\eea
Here $\tau_\epsilon$ is the Prym period matrix associated to the twist $\epsilon$.
In genus $h=2$, the even spin structures $\delta$ fall into two groups, depending
on whether $\delta+\epsilon$ is even or odd. The group with $\delta+\epsilon$ even
consists of 6 elements, which can be divided themselves into $\delta_i^+$ and
$\delta_j^+$, $j=2,3,4$, $\delta_j^-=\delta_j^++\epsilon$. These are the spin
structures occurring in the above formula for $d\mu_C[\delta;\epsilon,\alpha](p_L)$.
The Schottky relations imply that the choice of $j$ is immaterial.

\medskip
The asymptotic behavior of the measure
$d\mu[\delta;\epsilon,\alpha](p_L)$ is now easily determined along
the divisor of separating nodes. For example, in the sector $\epsilon=(0\, 0|0\, {1\over 2})$,
$\alpha=(0\, 0|{1\over 2}\, 0)$,
\bea
\sum_\delta
\<\alpha|\delta\>\Xi_6[\delta]\tet[\delta]^2
\tet[\delta+\epsilon]^2
\not\to 0,
\eea
so that the KKS cosmological constant does not vanish point by point on moduli space.

\newpage

\section{Outline of the Derivation}
\setcounter{equation}{0}

We provide now an outline of the construction of the
scattering amplitudes ${\bf A}_{II}$ described in Section \S 2.
In the Neveu-Schwarz-Ramond formulation of superstrings,
the superstring action is given by
\bea
I_m(x^\mu,\psi_\pm^\mu;g_{mn},\chi_m{}^\alpha)
&=&{1\over 4\pi}\int_\Sigma d^2z(\p_zx^\mu\p_{\bar z}x^\mu-
\psi_+^\mu\p_{\bar z}\psi_+^\mu
-\psi_-^\mu\p_z\psi_-^\mu
\nonumber\\
&&
\qquad\quad
+\chiz \psi_+^\mu\p_zx^\mu+
\chi_z{}^-\psi_-\p_{\bar z}x^\mu-{1\over 2}
\chiz\chi_z{}^-\psi_+^\mu\psi_-^\mu).
\eea
Here we have fixed a smooth surface $\Sigma$ of genus $h$,
$g_{mn}$ is a metric on $\Sigma$, and
$x^\mu$, $0\leq\mu\leq 9$, are scalar fields on $\Sigma$ which
can be interpreted geometrically as a map from $\Sigma$ into
10-dimensional flat Minkowski space-time.
The fields $\psi_\pm^\mu$ and
$\chi_m{}^\alpha$ are respectively
(anti-commuting) Majorana-Weyl spinors and gravitino fields,
defined with respect
to a given spin structure $\delta$, so that
$\psi_\pm\in \Lambda^{\pm{1\over 2},0}[\delta](\Sigma)$
and $\chi_{\bar z}{}^+\in \Lambda^{0,1}\otimes \Lambda^{-{1\over 2},0}[\delta](\Sigma)$,
$\chi_z{}^-\in \Lambda^{1,0}\otimes \Lambda^{0,-{1\over 2}}[\delta](\Sigma)$
if we view $\delta$ as a choice of a square root $\Lambda^{{1\over 2},0}[\delta](\Sigma)$
of the canonical bundle of $\Sigma$.

\medskip
The sums over the fluctuating worldsheets spanned by evolving strings
are realized by summing over all fields $x^\mu$, $g_{mn}$, $\psi_\pm^\mu$,
$\chi_m{}^\alpha$. Without the spinor fields $\psi_\pm^\mu$ and $\chi_m{}^\alpha$,
the action $I_m$ would reduce to the action for harmonic maps from
$\Sigma$ in flat space-time, and its conformal invariance would clearly
produce an integral over the moduli space ${\cal M}_h$ of Riemann surfaces
of genus $h$. In the present superstring context, the metric $g_{mn}$ has been replaced
by the ``supergeometry" $(g_{mn},\chi_m{}^\alpha)$, and the action acquires
a new symmetry, namely local supersymmetry.
We discuss geometric aspects of this symmetry before returning
to the evaluation of the sums over fluctuating worldsheets.

\subsection{Two-dimensional supergeometries and supermoduli}

The infinitesimal generator of a local supersymmetry is a spinor field $\delta\zeta^\alpha$,
and its infinitesimal action on supergeometries is
\bea
\delta e_m{}^a=\zeta \gamma^a\chi_m,
\quad \delta\chi_m{}^\alpha=-2\nabla_m\zeta^\alpha,
\eea
with similar actions on pairs $(x^\mu,\psi_\pm^\mu)$. Here $e_m{}^a$ is an orthonormal
frame for the metric $g_{mn}=e_m{}^ae_n{}^b\delta_{ab}$. There is an
evident similarity between
local supersymmetry transformations and infinitesimal diffeomorphisms, which
are generated by a vector field $\delta v^\alpha$, and are given by
$\delta e_m{}^a=
v^n\nabla_ne_m{}^a+e_n{}^a\nabla_m v^n$,
$\delta\chi_m{}^a=\delta v^n\nabla_n\chi_m{}^a+\chi_n{}^a\nabla_m\delta v^n$.
This similarity can be made more precise in the superspace formalism
\cite{superg}.
Let $s\Sigma$ be a supermanifold with $\Sigma$ as body, and local coordinates
${\bf z}=(z^M)=(z,\bar z,\theta,\bar\theta)$, where $\theta,\bar\theta$ are anti-commuting.
A supergeometry can then be identified with a superframe (or superzweibein) $E_M{}^A$
and a $U(1)$ superconnection $\Omega_M$ satisfying the Wess-Zumino torsion
constraints
\bea
T_{ab}{}^c=T_{\alpha\beta}{}^\gamma=0,
\qquad T_{\alpha\beta}{}^c=2(\gamma^c)_{\alpha\beta},
\eea
where the torsion $T_{AB}{}^C$ and curvature $R_{AB}$ of the superconnection
$\Omega_M$ are defined by
$[{\cal D}_A,{\cal D}_B]=T_{AB}{}^C{\cal D}_C+in R_{AB}$,
and ${\cal D}_AV=E_A{}^M(\p_M V_B+in\Omega_M V)$ is the covariant derivative on
fields $V$ of U(1) weight $n$. The group $sDiff(\Sigma)$ acts on
supergeometries by
\bea
\delta E_M{}^B
=
E_M{}^A({\cal D}_A\delta V^B-\delta V^CT_{CA}{}^B+\delta V^C\Omega_CE_A{}^B).
\eea
The equivalence with the earlier definition of a supergeometry as $(g_{ab},\chi_a{}^\alpha)$
is obtained by putting the superframe $E_M{}^A$ in the Wess-Zumino gauge, where the
frame components $E_{\mu}{}^\alpha$ and $E_{\mu}{}^a$ are required to
satisfy $E_\mu{}^\alpha\sim\delta_\mu{}^\alpha+\theta^\nu e_{\nu\mu}^{*\alpha}$,
$E_\mu{}^a\sim \theta^\nu e_{\nu\mu}^{**a}$ for some
$e_{\nu\mu}^{*a}$ and $e_{\nu\mu}^{**a}$ symmetric in $\nu$ and $\mu$.
In such a gauge, the component $E_m{}^a$ takes the form
\bea
E_m{}^a=e_m{}^a+\theta\gamma^a\chi_m-{i\over 2}\theta\bar\theta e_m{}^aA,
\eea
with all other components of $E_M{}^A$ and $\Omega_M$
expressible as well in terms of $e_m{}^a$, $\chi_m$, and $A$.
The auxiliary field $A$ can be set to 0 for all practical purposes, and we obtain
in this manner the desired identification of the supergeometry $E_M{}^A,\Omega_M$ with
the pair $e_m{}^a,\chi_m$. A vector field $\delta V^M$ in superspace
can then be decomposed into components $\delta v^m$
and $\delta\zeta^\alpha$, and the corresponding superdiffeomorphisms decompose
correspondingly into diffeomorphisms and local supersymmetry
transformations. Similarly, super Weyl transformations
can be defined which decompose into the standard Weyl transformations
and the super Weyl transformations proper.
The fields $x^\mu$ and $\psi_\pm^\mu$ can also be grouped
into a scalar superfield $X^\mu(z,\theta,\bar\theta)
=x^\mu+\theta\psi_+^\mu+\bar\theta\psi_-^\mu$.
In Wess-Zumino gauge, the covariant
derivative of a superfield $V(z,\theta,\bar\theta)
=V_0+\theta V_++\bar\theta V_-$ of U(1) weight $n$ becomes
\cite{dp89},
\bea
\label{calD}
{\cal D}_-^{(n)}V
=
V_-+\bar\theta(\p_{\bar z}V_0+{1\over 2}\chi_{\bar z}{}^+V_+)
-
\theta\bar\theta(\p_{\bar z}V_++{1\over 2}\chi_{\bar z}{}^+\p_zV_0
+n\p_z\chi_{\bar z}{}^+\,V_0-{1\over 4}\chiz\chi_z{}^-V_-)
\eea
Introducing the measure $d^{2|2}{\bf z}=d^2z\,d\theta d\bar\theta$
and the volume element $E({\bf z})={\rm sdet}\,E_M{}^A
=({\rm det}\,e_m{}^a)\,(1+{1\over 4}\theta\bar\theta\chiz\chi_z{}^-)$,
the action $I_m$ can be expressed in the following
manifestly supersymmetric and super Weyl invariant form
\bea
I_m(E_M{}^A,X^\mu)={1\over 4\pi}\int
d^{2|2}{\bf z}\,E({\bf z},\bar{\bf z})\,{\cal D}_+X^\mu{\cal D}_-X^\mu.
\eea

\medskip

$\bullet$ Associated to each supergeometry is a notion of superholomorphicity.
In the superspace formalism, we can define a supercomplex structure $J_M{}^N$ by
\cite{dp88}
\bea
J_M{}^N=E_M{}^a\epsilon_a{}^bE_b{}^N+E_M{}^\alpha(\gamma_5)_\alpha{}^\beta E_\beta{}^N
\eea
which satisfies $J_M{}^NJ_N{}^P=-\delta_M{}^N$ and the integrability condition
$d\zeta^M\equiv 0\ ({\rm mod}\ \zeta^N)$, where $\zeta^M \equiv dz^M-idz^NJ_N{}^M$.
A scalar function $f(z,\theta)$ is defined then to be superholomorphic if
$J_M{}^N{\cal D}_Nf=0$, or equivalently ${\cal D}_-f=0$. More generally, a field
$\hat\omega(z,\theta)$ on $s{\cal M}$ of U(1) weight $n$ is said to be superholomorphic if
\bea
{\cal D}_-^{(n)}\hat\omega=0,
\eea
where ${\cal D}_-^{(n)}$ is the covariant derivative on fields of weight $n$ with
respect to the given supergeometry. In particular,
for a form $\hat\omega$ of U(1) weight $1/2$ of the form $\hat\omega(z,\theta)$ as
$\hat\omega(z,\theta)=\omega_0+\theta\omega_+$, the superholomorphicity condition
is equivalent to the following
system of partial differential equations on $\Sigma$
\bea
\p_{\bar z}\omega_0+{1\over 2}\chi_{\bar z}{}^+\omega_+=0,
\qquad
\p_{\bar z}\omega_+
+{1\over 2}\p_z(\chi_{\bar z}{}^+\,\omega_0)=0.
\eea

\medskip
$\bullet$ A key property of supergeometries $(g_{mn},\chi_m{}^\alpha)$
defined by an even spin structure $\delta$
is that, generically, there exists a unique basis of superholomorphic forms $\hat\omega_I$
of U(1) weight $1/2$ dual to the $A_I$ cycles, and hence a {\it super period matrix}
$\hat\Omega_{IJ}$ can be defined by
\bea
\oint_{A_J}\hat\omega_J=\delta_{IJ},
\quad\oint_{B_J}\hat\omega_J=\hat\Omega_{IJ}.
\eea
Here the integral over a cycle $C$ of a form $\hat\omega=\omega_0+\theta\omega_+$ of
U(1) weight 1/2 is
defined by $\oint_C\hat\omega= \oint_C (dz\,\omega_+-{1\over 2}d\bar z
\chi_{\bar z}{}^+\omega_0)$. Explicitly, $\hat\Omega_{IJ}$ and $\Omega_{IJ}$
can be determined from each other by the following equation
\bea
\label{omegahatomega}
\hat\Omega_{IJ}=\Omega_{IJ}-{i\over 8\pi}
\int\int d^2y\, d^2x\ \omega_I(x)\chi_{\bar x}{}^+\hat S_\delta(x,y)\chi_{\bar y}{}^+\omega_J(y),
\eea
where $\omega_I$ is a basis of holomorphic 1-forms with respect to the complex structure
defined by $g_{mn}$, and $\hat S_\delta(x,y)$ is the modification of the Szeg\"o kernel
of $g_{mn}$ by
\bea
\hat S_\delta(z,w)
=
S_\delta(z,w)-{i\over 16\pi^2}
\int\int d^2u\, d^2v\ S_\delta(z,u) \chi_{\bar u}{}^+\p_u\p_v\ln E(u,v)
\chi_{\bar v}{}^+\hat S_\delta(v,w),
\eea
with $E(u,v)$ the prime form.
By construction, the super period matrix $\hat\Omega_{IJ}$
is invariant under all symmetry transformations including supersymmetry.
In genus $h=2$, the super period matrix $\hat\Omega_{IJ}$ is always
well-defined for even spin structures.

\medskip
$\bullet$ We come now to the essential relation between superholomorphicity
and holomorphicity which underlies our derivation of superstring
scattering amplitudes. First, we note that there can be no
intrinsic relation between the
superholomorphicity
of a form $\hat\omega=\omega_0+\theta\omega_+$ and the holomorphicity of
its components, if the latter notion of holomorphicity is taken with
respect to the metric $g_{mn}$. This is simply because the conformal class of the
metric $g_{mn}$ is not left invariant under supersymmetry transformations.
The only candidate for a supersymmetric substitute is the super period matrix
$\hat\Omega_{IJ}$.

Thus, we choose a metric $\hat g_{mn}$ whose period matrix
is $\hat\Omega_{IJ}$. Such a metric is only determined up to diffeomorphisms,
and the relation we need between superholomorphicity and holomorphicity
with respect to the metric $\hat g_{mn}$ has to take into account this
gauge choice. Furthermore, because of the deformation of complex
structure from $g_{mn}$ to $\hat g_{mn}$, the forms $\omega_0$ and
$\omega_+$ are no longer pure $(p,0)$-forms with respect to $\hat g_{mn}$,
so they cannot possibly be holomorphic. The guiding principle is that
the $\theta$-component of a superholomorphic form with respect to
the supergeometry $(g_{mn},\chi_m{}^\alpha)$ is a holomorphic form
with respect to $\hat g_{mn}$, up to a de Rham exact differential.
We provide below some explicit examples of this relation between
holomorphicity and superholomorphicity
in the case of genus $h=2$. In this case, the
calculations are relatively simpler because
there are only 2 odd supermoduli $\zeta^\alpha$, and perturbation theory
need only be worked out to first even order $\zeta^1\zeta^2$. Similar
formulas can be expected to hold in higher genus.

(a) Let $\hat\omega(z,\theta)$ be a weight 1/2 superholomorphic form
with respect to $(g_{mn},\chi_m{}^\alpha)$. Then
\bea
\int d\theta\,\hat\omega=\omega(z)+d\lambda(z),
\eea
where $\omega(z)$ is a holomorphic $(1,0)$-form with respect to $\hat g_{mn}$,
and $\lambda(z)$ is a $C^\infty$ scalar function. Under changes of metrics
$\hat g_{mn}$, $\lambda$ changes by $\delta\lambda=-\delta v^z\omega(z)$.

(b) Let ${\cal E}_\delta({\bf z},{\bf w})$ be the super prime form
(see \cite{dp89} for the definition). Then there exists a scalar
function $\hat f_0(z,w)$ so that
\bea
&&\int d\theta_{z_i}\int d\theta_{z_j}
\,E({\bf z}_i)\,E({\bf z}_j)\
{\cal D}_+^{{\bf z}_i}{\cal D}_+^{{\bf z}_j}\ln {\cal E}_\delta({\bf z}_i,
{\bf z}_j)
\nonumber\\
&&
\qquad\qquad
=
dz_i\wedge dz_j \,\p_{z_i}\p_{z_j}\ln E(z_i,z_j)
-d_id_j\hat f_0(z_i,z_j),
\eea
up to Dirac measures supported at coincident points. By the cancelled propagator argument,
amounting to an analytic continuation in $s_{ij}=-2k_i\cdot k_j$
\cite{dp92}, such Dirac measures can always be dropped
in presence of the factor $\prod_{i<j}E(z_i,z_j)^{k_i\cdot k_j}$.
Thus, up to exact de Rham differentials, the highly non-holomorphic
term ${\cal D}_+^{\bf z}{\cal D}_+^{\bf w}\ln {\cal E}_\delta({\bf z},{\bf w})$
reduces to the holomorphic function $\p_z\p_w\ln E(z,w)$.

(c) The relation between holomorphicity and superholomorphicity
leads to many new holomorphic forms on moduli space, the existence of which
may not have been suspected otherwise. For example, if we write $\hat\omega_I
=\omega_{I0}+\theta\omega_{I+}$, and let $\lambda_I$ be the scalar function
defined up to a constant by $\hat\omega_{I0}=\omega_I(z)+d\lambda_I(z)$,
then the expression
\bea
\Pi_{IJ}^{(1)}(z)
=
\omega_I(z)\lambda_J(z)-\omega_J(z)\lambda_I(z)-
\hat\omega_{I0}(z)\hat\omega_{J0}(z)
\eea
is a holomorphic form. Many other holomorphic forms in more variables can be
constructed in the same manner from components of superholomorphic forms.

(d) In superstring perturbation theory, it is necessary to consider
superholomorphic forms with certain non-trivial monodromies, as in
(\ref{monodromy}). Here the relation between superholomorphicity and
holomorphicity has been established so far only through involved
explicit calculations, for the specific superholomorphic forms
arising from correlation functions of scalar superfields. The
relation between the holomorphic form ${\cal H}$ and the combination
$\sum_\delta\epsilon_\delta{\cal B}[\delta]$ described in Section \S
2 is a prime example.

\bigskip

$\bullet$ The supermoduli space of the surface $\Sigma$ is defined to be
\bea
s{\cal M}_h=\{(g_{mn},\chi_m{}^\alpha)\}/\{{\rm symmetries} \}
\eea
where the symmetries are generated by Weyl, super Weyl, diffeomorphisms,
and supersymmetry transformations.
The tangent space $T(s{\cal M}_h)$
to $s{\cal M}_h$ decomposes as
$\{\delta g_{mn}\}\oplus\{\delta\chi_m{}^\alpha\}$. In
local complex coordinates $z,\bar z$
for the metric $g_{mn}$, we may set $\delta g_{\bar zz}=0$ and $\delta\chi_{\bar z}{}^-
=\delta\chi_z{}^+=0$ by Weyl and super Weyl transformations. The dimension of
the remaining modes $\delta g_{zz}$ and $\delta\chi_{\bar z}{}^+$
in $T(s{\cal M}_h)$ after diffeomorphisms and supersymmetry transformations can be
easily determined by their values at $\chi=0$, where they are given respectively
by the codimensions of the $\bar\partial$
operators on tensors of $U(1)$ weights 2 and $3/2$ respectively.
By the Riemann-Roch theorem, we obtain
\bea
{\rm dim}\,(s{\cal M}_h)
=
\cases{(0|0), & if $h=0$\cr (1|0)_e\ {\rm or}\ (1|1)_0, & if $h=1$\cr
(3h-3|2h-2), & if $h\geq 2$,\cr}
\eea
where the dimensions indicated for genus $1$ depend on whether the spin structure
$\delta$ is even or odd, as indicated by the indices $e$ or $o$.

\subsection{Functional integrals}

We return to the derivation of the superstring scattering
amplitudes. We start from sums over fluctuating worldsheets given by
the following functional integrals
\bea
{\bf A}[\delta]
=
\int DE_M{}^AD\Omega_M\delta(T)
\int \prod_{i=1}^Nd^{2|2}{\bf z}_i E({\bf z}_i)
\int DX^\mu e^{-I_m}\prod_{i=1}^NV({\bf z}_i,\bar{\bf z}_i;\epsilon_i,\bar\epsilon_i,k_i)
\eea
where $V({\bf z}_i,\bar{\bf z}_i;\epsilon_i,\bar\epsilon_i,k_i)
={\rm exp}(ik_i^\mu X^\mu({\bf z}_i)+\epsilon_i^\mu{\cal D}_+X^\mu+\bar\epsilon_i^\mu{\cal D}_-X^\mu)$,
$k^2=k\cdot\epsilon=k\cdot\bar\epsilon=0$, is the generating vertex for the graviton
multiplet \cite{D'Hoker:1987bh}.
Factoring out all symmetries reduces these functional integrals to an
integral over supermoduli space (\cite{dp88}, eq. (3.143))
\bea
{\bf A}[\delta]
&=&
\int |\prod_A dm^A|^2
\int \prod_{i=1}^Nd^{2|2}{\bf z}_i E({\bf z}_i)
\int D(B\bar BC\bar C X^\mu)e^{-I_m-I_{gh}}
\nonumber\\
&&
\qquad\qquad\qquad
\times\,
|\prod_A\delta(\<H_A|B\>)|^2
V({\bf z}_i,\bar{\bf z}_i;\epsilon_i,\bar\epsilon_i,k_i)
\eea
Here $m^A$ are $(3h-3|2h-2)$ local complex parameters for a slice $\hat{\cal S}$
for supermoduli space,
\bea
\label{HA}
(H_A)_-{}^z=(-)^{A(M+1)}E_-{}^M{\p E_M{}^z\over \p m^A}
\eea
are the super Beltrami differentials tangent to the
gauge slice $\hat{\cal S}$, and the Faddeev-Popov determinants of the gauge-fixing
procedure have been encoded in an integration over the superghost fields
$B=\beta+\theta b$, $C=c+\theta\gamma$
of $U(1)$ weights $3/2$ and $-1$ respectively
with action $I_{gh}={1\over 2\pi}\int d^{2|2}{\bf z}\,E
(B{\cal D}_-C+\bar B{\cal D}_+\bar C)$. In components, the superghost
action can be expressed as
\bea
I_{gh}=\int d^2z \,\{b\p_{\bar z}c+\beta\p_{\bar z}\gamma+
\chiz S_{gh}
+c.c.\},
\eea
where $S_{gh}={1\over 2}b\gamma-{3\over 2}\beta\p_zc-(\p_z\beta)c$ is the ghost
supercurrent.

\medskip
$\bullet$
The integrals ${\bf A}[\delta]$ are only a preliminary step in constructing the superstring
scattering amplitudes. To obtain these,
one has to identify in ${\bf A}[\delta]$
the contributions of each chiral sector, and sum these contributions
over $\delta$,
with suitable phases $\epsilon_\delta$ so as to insure modular invariance.
The chiral sector corresponds to the correct degrees of freedom of the Minkowski
formalism, and the summation over spin structures is the Gliozzi-Scherk-Olive
projection, necessary for eliminating tachyons and insuring space-time supersymmetry.

The identification of the contributions of each chiral sector is provided
by the chiral splitting theorem of \cite{dp89}, which asserts that
\bea
\int DX^\mu
\prod_{i=1}^Ne^{-I_m}
V({\bf z}_i,\bar{\bf z}_i;\epsilon_i,\bar\epsilon_i,k_i)
=
\int dp_I^\mu
\big|\<Q(p_I){\rm exp}({1\over 2\pi}\int \chi S_m)
\prod_{i=1}^NW({\bf z}_i;\epsilon_i,k_i)\>_+\big|^2
\eea
where $Q(p_I)={\rm exp}\{ip_I^\mu\oint_{B_I}dz\,\p_zx_+^\mu(z)\}$, $W({\bf z};\epsilon,k)$
is the chiral generating vertex given by
\bea
W({\bf z};\epsilon,k)
=
{\rm exp}\{ik^\mu(x_+^\mu+\theta\psi_+^\mu)(z)
+
\epsilon^\mu(\psi_++\theta\p_zx_+^\mu)(z)\}.
\eea
The expectation value $\<\cdot\>_+$ is taken with respect to
an effective bosonic chiral field $x_+^\mu(z)$ with propagator
$\<x_+^\mu(z)x_+^\nu(w)\>=-\delta^{\mu\nu}\ln E(z,w)$, and a fermionic
field $\psi_+^\mu(z)$ with propagator
$\<\psi_+^\mu(z)\psi_+^\nu(w)\>=-
\delta^{\mu\nu}S_\delta(z,w)$, where $S_\delta(z,w)$
is the Szeg\"o kernel. The expression $S_m$
is the effective matter supercurrent $S_m=-{1\over 2}\psi_+^\mu\p_zx_+^\mu$.
The point of this formula is that,
by introducing the parameters $p_I^\mu$,
the real bosonic field $x^\mu(z)$
has been replaced by a chiral field $x_+^\mu$, and that all terms mixing opposite
chiralities such as $\chi_{\bar z}{}^+,
\psi_+^\mu$ with $\chi_z{}^-,\psi_-^\mu$ have cancelled out.
Physically, as in the case of the bosonic string
discussed in \cite{vv87}, the parameters $p_I^\mu$, $1\leq I\leq h$, $0\leq \mu\leq 9$,
can be interpreted as internal loop momenta.

\medskip

$\bullet$
We also need to split chirally the super volume form $d^{2|2}{\bf z}E({\bf z},\bar{\bf z})$ on
the superworldsheet. From \cite{dp88}, eqs. (3.32)-(3.33), we have
$d^{2|2}{\bf z}\,E({\bf z},\bar {\bf z})= d\bar\theta \wedge e^{\bar z}\wedge d\theta\wedge e^z$
with $e^z=dz-{1\over 2}\theta\chiz d\bar z$. If we let now
\bea
{\cal V}(z;\epsilon,k)
=
\int d\theta e^zW({\bf z};\epsilon,k)
=\epsilon^\mu\{(\p_zx_+^\mu-ik^\nu\psi_+^\mu\psi_+^\nu)dz
-
{1\over 2}d\bar z\chiz\psi_+^\mu\}e^{ik\cdot x_+(z)},
\eea
we can write
\bea
\label{gaugefixed3}
{\bf A}[\delta]
=
\int dp_I^\mu\int_{{\cal S}}
\bigg|
\prod_{A=1}^{(3h-3|2h-2)} dm^A \,\<\prod_A\delta(H_A|B)\,
Q(p_I){\rm exp}\{{1\over 2\pi}\int \chi S\}\prod_{j=1}^N{\cal V}_j\>
\bigg|^2
\eea
Here the expectation value is with respect to all chiral fields $x_+^\mu,\psi_+^\mu$,
$b,c,\beta,\gamma$, and $S=S_m+S_{gh}$ is the total supercurrent, incorporating
the effective matter supercurrent $S_m$ as well as the ghost supercurrent
$S_{gh}$ from the superghost action $I_{gh}$. Naively, after implementation
of the Gliozzi-Scherk-Olive
projection, the scattering amplitude ${\bf A}_{II}$ should be given by
\bea
{\bf A}_{II}
=
\int dp_I^\mu\int_{{\cal S}}
\bigg|
\prod_{A=1}^{(3h-3|2h-2)} dm^A \,\sum_\delta \epsilon_\delta\<\prod_A\delta(H_A|B)\,
Q(p_I){\rm exp}\{{1\over 2\pi}\int \chi S\}\prod_{j=1}^N{\cal V}_j\>
\bigg|^2
\eea
We should stress that all complex coordinates and correlation functions are at this time
written with respect to the metric $g_{mn}$ from the slice $\hat{\cal S}$.

\medskip
So far, the gauge-fixed formula (\ref{gaugefixed3}) holds for an arbitrary choice
of $(3h-3|2h-2)$-dimensional slice $\hat{\cal S}$ in the space of supergeometries.
The issue is whether the odd supermoduli $dm^\alpha$ can be integrated out
to produce forms a global form over moduli space.
Perhaps surprisingly, this turns out not to be the case with the naive
projections used in the early 1980's, and the origin
of the problem has been somewhat of a mystery ever since \cite{vv87, ambiguities}.
We discuss it and its resolution in the next section.

\subsection{Deformation of complex structures}

The above Faddeev-Popov type gauge-fixing procedure shows that, upon cancellation of
all anomalies, the sums over all supergeometries can be reduced to sums over
supermoduli space, after factoring out all symmetries. The new difficulty
peculiar to superstring perturbation theory is that the superstring amplitudes
have to be expressed as sums over moduli space and not as sums over supermoduli
space. To go from supermoduli to moduli, a correct structure for
supermoduli space as a fibration over moduli space has to be identified, and the
odd supermoduli degrees of freedom integrated out. This deceptively simple problem
has to be approached with some care.

\medskip
$\bullet$ The projection $(g_{mn},\chi_m{}^\alpha)\to g_{mn}$ from supergeometries to geometries
seems a natural candidate for constructing such a fibration.
However, it is not well-defined as a projection from
supermoduli space to moduli space, as supergeometries equivalent under supersymmetries
do not project to geometries equivalent under diffeomorphisms and Weyl transformations
\bea
\matrix{(g_{mn},\chi_m{}^\alpha)&\sim & (g_{mn}+\delta g_{mn},\chi_m{}^\alpha
+\delta \chi_m{}^\alpha)\cr
\downarrow & &\downarrow\cr
g_{mn}&\not\sim  & g_{mn}+\delta g_{mn}\cr}
\eea
The only alternative is to
rely instead on the super period matrix $\hat\Omega$ and the projection
\bea
\matrix{(g_{mn},\chi_m{}^\alpha)\cr\downarrow\cr\hat\Omega_{IJ}\cr}
\eea
which is invariant under supersymmetry and does descend to the complement of a lower-dimensional
subvariety in supermoduli space. We develop now the gauge-fixing procedure
based on this projection.

\medskip

$\bullet$ As in our earlier discussion of the relation between superholomorphicity
and holomorphicity with respect to $\hat\Omega_{IJ}$,
the projection $(g_{mn},\chi_m{}^\alpha)\to \hat\Omega_{IJ}$
has to be supplemented by a choice of metric $\hat g_{mn}$ whose period matrix
is $\hat\Omega_{IJ}$. There is no canonical $\hat g_{mn}$,
and different choices of $\hat g_{mn}$ are related infinitesimally by
$\delta\hat g_{mn}=\nabla_m \delta v_n+\nabla_n\delta v_n$,
where $\delta v^n$ is a smooth vector field on $\Sigma$. In genus $h=2$,
the deformation from $\hat g_{mn}$ to $g_{mn}$ is only of first order in $\zeta^1\zeta^2$,
and we may define its Beltrami differential
$\hat \mu_{\bar z}{}^z={1\over 2}\hat g^{z\bar z}
g_{\bar z\bar z}$ in local holomorphic coordinates for $\hat g_{mn}$.
Then $\hat \mu_{\bar z}{}^z$ is defined by the condition
\bea
i\int_\Sigma \omega_I\omega_J\hat \mu_{\bar z}{}^z=\Omega_{IJ}-\hat\Omega_{IJ}.
\eea
This equation determines $\hat \mu_{\bar z}{}^z$ only up to a gauge choice of
$\delta \hat \mu_{\bar z}{}^z=\p_{\bar z}\delta v^z$.

A choice of metrics is necessary because the correlation functions of conformal and
superconformal field theories require an underlying geometry or supergeometry,
and not just an equivalence class under diffeomorphisms and/or supersymmetry
transformations. It will be an important check of the consistency of
our gauge-fixing procedure for superstring amplitudes that, after integration
over all insertion points, the final amplitude is independent of the choice
of $\hat \mu_{\bar z}{}^z$.

\medskip

$\bullet$ We can construct a slice $\hat{\cal S}$ for supermoduli space which
fibers over the period matrices $\hat\Omega$ as follows.
Let $\hat\Omega_{IJ}$, $1\leq I\leq J\leq 2$,
be the 3 local holomorphic coordinates for moduli space, and choose a 3-dimensional
slice $\hat S$ of frames $\hat e_m{}^a$ whose period matrices are the matrices
$\hat\Omega_{IJ}$. For each of these frames $\hat e_m{}^a$, choose 2 generic
gravitino sections $\hat\chi_{\alpha}$, $\alpha=1,2$, and set $\hat\chi
=\sum_{\alpha=1}^2\zeta^\alpha\hat\chi_\alpha$, where $\zeta^\alpha$
are 2 anticommuting parameters. We can choose next a $(3|2)$-dimensional slice
of supergeometries $(e_m{}^a,\chi)$ whose period matrices $\Omega_{IJ}$ and
$\hat\Omega_{IJ}$ satisfy the equation (\ref{omegahatomega}). This can clearly
be done, because $\Omega_{IJ}$ and $\hat\Omega_{IJ}$
differ by terms of order $O(\zeta^1\zeta^2)$, and thus gravitino sections $\hat\chi_\alpha$
with respect to
$\hat e_m{}^a$ can be considered as gravitino sections $\chi_\alpha$ with respect
to $e_m{}^a$.

There are three significant complications in this gauge-fixing procedure, compared
to the earlier one based on the simpler but ill-behaved projection
$(g_{mn},\chi_m{}^\alpha)\to g_{mn}$:

(a) The first is that
the Beltrami superdifferentials $H_A=\bar\theta(\hat \mu_A-\theta\chi_A)$
defined by the slice $\hat{\cal S}$ have components $\hat \mu_A$ and
$\nu_A$ which are both non-vanishing, unlike in the earlier case
where one of the components $\hat \mu_A$ or
$\nu_A$ is always 0. This reflects the fact that, to maintain $\hat\Omega_{IJ}$
fixed, both $g_{mn}$ and $\chi_m{}^\alpha$ have to be deformed
simultaneously.

(b) The second is that
the correlation functions of the underlying conformal field theories
are expressed in the background of the metric $g_{mn}$. To re-express them in
the background of the metric $\hat g_{mn}$, we need to carry out a deformation of
complex structures, and hence an insertion of the stress tensor $T(z)$.

(c) The third is that the vertex operators ${\cal V}$ have to be
deformed as well. This produces new vertex operators \bea {\cal
V}(z)&=&{\cal V}^{(0)}(z)+{\cal V}^{(1)}(z)+{\cal V}^{(2)}(z) \eea
where ${\cal V}^{(0)}$ is the naive vertex operator of
(\ref{vertex}), and ${\cal V}^{(1)}$, ${\cal V}^{(2)}$ are
deformation corrections.

\medskip

Taking all these points into account,
we obtain the following first formula for the gauge-fixed amplitude,
\bea
{\bf A}[\delta]
=\int dp_I^\mu \int_{{\cal M}_2\times\Sigma^N}
\bigg|\ {\cal B}[\delta](z_i;k_i,\epsilon _i ;p_I^\mu)\,\bigg|^2
\eea
where the chiral forms ${\cal B}[\delta](z_i;k_i, \epsilon_i ;p_I^\mu)$ are given by
\bea
{\cal B}[\delta](z_i;k_i, \epsilon_i ;p_I^\mu)
&=&\prod_{I\leq J}d\hat\Omega_{IJ}\int\prod_{\alpha=1,2}d\zeta^\alpha
{\<\prod_a b(p_a)\prod_{\alpha}\delta(\beta(q_\alpha))\>
 \over {\rm det}\,\Phi_{IJ+}(p_a)\cdot {\rm det}\,\<H_\alpha|\Phi_\beta^*\>}
\nonumber\\
&&
\qquad\quad
\times
\<Q(p_I){\rm exp}\{{1\over 2\pi}
\int (\chi(z) S(z)+ \hat \mu(z) T(z))\}\prod_{j=1}^N{\cal V}_j\>.
\eea
Here $p_a$, $1\leq a\leq 3$, and $q_\alpha$, $1\leq\alpha\leq 2$,
are two sets of arbitrary auxiliary points on $\Sigma$,
$\Phi_{IJ}(z,\theta)=\Phi_{IJ0}+\theta\Phi_{IJ+}$ is the basis
of superholomorphic differentials of weight $3/2$ defined by
$-{i\over 2}(\hat\omega_I{\cal D}_+\hat\omega_J+\hat\omega_I{\cal D}_+\hat\omega_J)$,
$\Phi_\beta^*=\Phi_{\beta0}+\theta\Phi_{\beta+}^*$ is another basis
of superholomorphic differentials of weight 3/2 normalized by
$\Phi_{\beta0}^*(q_\alpha)=\delta_{\alpha\beta}$ and $\Phi_{\beta+}^*(p_a)=0$,
and $S(z)$ is the supercurrent.

\medskip

The three points which we stressed above
are reflected in the appearance in this formula of the full vertex operator
${\cal V}(z_i)$, of the finite-dimensional determinants
$\Phi_{IJ+}(p_a)$ and $\<H_\alpha|\Phi_\beta^*\>$
corresponding to the
gauge slice constructed, and of the insertion of the stress tensor $T_{zz}$
implementing the deformation of complex structures.

\medskip
At this moment, after the deformation of complex structures has been carried now,
all correlation functions in ${\cal B}[\delta](z_i;k_i,\epsilon_i;p_I^\mu)$
are expressed with respect to the metric $\hat g_{mn}$. The metric $g_{mn}$ and
its period matrix no longer enter the picture, and we can now just denote
$\hat\Omega_{IJ}$ by $\Omega_{IJ}$ for simplicity.

\subsection{Gauge slice independence}

We derive next explicit formulas for
${\cal B}[\delta](z_i;k_i, \epsilon_i;p_I^\mu)$.
In the process, we establish the independence of the amplitudes from all choices
entering the gauge-fixing process. This also facilitates later
explicit evaluations, since we shall be free to make convenient choices
for different calculations.

\medskip
$\bullet$ We begin with the $0$-point function, in which case
the dependence of ${\cal B}[\delta]$ on the external momenta
is trivial, and ${\cal B}[\delta]$ reduces essentially
to the measure $d\mu_2[\delta]$,
\bea
{\cal B}[\delta](p_I^\mu)
={\rm exp}(i\pi p_I^\mu\Omega_{IJ}p_J^\mu)
\ d\mu_2[\delta](\Omega),
\qquad\quad N=0,
\eea
and $d\mu_2[\delta]$ is itself given by
\bea
d\mu_2[\delta](\Omega)
&=&
{\<\prod_a b(p_a)\prod_{\alpha}\delta(\beta(q_\alpha))\>
 \over {\rm det}\,\Phi_{IJ+}(p_a)\cdot {\rm det}\,\<H_\alpha|\Phi_\beta^*\>}
\{1-{1\over 8\pi^2}\int
d^2z\chiz\int d^2w\chiw \<S(z)S(w)\>
\nonumber\\
&& \qquad\qquad\qquad\qquad\qquad\qquad +{1\over 2}\int
d^2z \hat \mu_{\bar z}{}^z\<T(z)\>\}
\eea
The correlation functions of the
supercurrent and stress tensor can now be evaluated in terms of
prime forms and Green's functions. This gives the expression
(2.8)-(2.12) for $d\mu_2[\delta]$ described earlier. The
independence of gauge choices is then obtained by showing that the
variations of $d\mu_2[\delta]$ under changes of $\chi_\alpha(z)$ as
well as $\mu(z)$ vanish point by point on the moduli space ${\cal
M}_2$. The case of $\chi_\alpha(z)=\delta(z,x_\alpha)$ is
particularly convenient. In this case, $d\mu_2[\delta]$ reduces to
the expression (\ref{mu2dirac}). This expression can be independently
verified to be independent of all points $p_a$, $q_\alpha$, and
$x_\alpha$. Note that the Beltrami differential $\hat \mu(z)$ has
cancelled out, so we have manifest independence from the choice of
metrics $\hat g_{mn}$.

\medskip

$\bullet$ Next, we show the gauge slice independence of the $N$-point function.
Since $d\mu_2[\delta]$ has been shown to be gauge slice independent,
and since the factor $\<Q(p_I)\prod_{i=1}^N{\cal V}(z_i)\>$ does not depend
on any gauge choice, the term ${\cal B}[\delta]^{(d)}$ is gauge-slice
independent. As for
the term ${\cal B}^{(c)}$,
we can show that it transforms as,
\bea
\delta {\cal B}[\delta]^{(c)}
=\sum_{i=1}^Nd_i{\cal R}_i[\delta]
\eea
with ${\cal R}_i[\delta]$ given respectively by
\bea
{\cal R}_i[\delta]&=&-\delta v^{z_i}\<Q(p_I)\prod_{j=1}^N{\cal V}_j^{(0)}\>\,d\mu_0[\delta]
\\
{\cal R}_i[\delta]
&=&
-\<Q(p_I)\delta\xi^+(z_i)\epsilon_i^\mu\psi_+(z_i)
e^{ik_ix_+(z_i)}({1\over 2\pi}\int \chi S\prod {\cal V}_l^{(0)}+
\sum_{j\not-i}{\cal V}_j^{(1)}\prod_{l\not=i,j}{\cal V}^{(0)})\>
d\mu_0[\delta]
\nonumber
\eea
under changes of Beltrami differentials by
$\delta \hat \mu_{\bar z}{}^z=\p_{\bar z}\delta v^z$ and changes of
gravitino slices by $\delta\chi_{\bar z}{}^+
=-
2\p_{\bar z}\delta\xi^+$, $\delta \hat \mu_{\bar z}=\delta\xi^+\chi_{\bar z}^+$.
Note that there are no
exterior derivative in the moduli variables $\Omega_{IJ}$. As explained in
Section \S 2, changes in ${\cal B}[\delta]$ of the above form leave
the integrated amplitudes invariant.

\subsection{Modular forms and $\tet$ constants}

The chiral amplitudes ${\cal B}[\delta](z_i;k_i,\epsilon_i;p_I^\mu)$ have now to
be evaluated, certainly explicitly enough so that the relative
phases $\epsilon_\delta$ can be determined which would lead to
a modular invariant integral formula for the superstring amplitude
${\bf A}_{II}(k_i,\epsilon_i)$. In principle, all the correlation functions
needed are of free fields, and the chiral determinants needed
can be obtained from the chiral bosonization formulas of
\cite{vv87-2}, \cite{fay}, \cite{faltings}. However, these formulas
depend typically on many extraneous points whose presence makes the
modular transformations obscure. For our purposes, it is then important
to eliminate completely these points, and remarkably, this turns out
to be possible.

\medskip

$\bullet$ We begin with the evaluation of the $0$-point function, or
equivalently, of $d\mu_2[\delta]$. Here we exploit the independence
of the expression (\ref{mu2dirac}) to work in the split gauge, where
the points $q_1,q_2$ are chosen to satisfy the $\delta$ dependent
relation
\bea
S_\delta(q_1,q_2)=0.
\eea
All dependence on $p_a$,
$q_\alpha$ then manifestly cancels out, and we obtain the expression
\bea
d\mu_2[\delta]
=
\prod_{I\leq J}d\Omega_{IJ}\tet[\delta]^4 \
{\<\nu_1|\nu_2\>{\cal M}_{\nu_1\nu_2}+\<\nu_2|\nu_3\>{\cal
M}_{\nu_2\nu_3}+ \<\nu_3|\nu_1\>{\cal M}_{\nu_3\nu_1}
\over
16\pi^2{\cal M}_{\nu_1\nu_2}^2{\cal M}_{\nu_2\nu_3}^2 {\cal
M}_{\nu_3\nu_1}^2}
\eea
where the bilinear $\tet$-constant ${\cal
M}_{\nu_i\nu_j}$ is defined by
\bea {\cal M}_{\nu_i\nu_j} =
\p_1\tet[\nu_i]\p_2\tet[\nu_j]-\p_2\tet[\nu_i]\p_1\tet[\nu_j].
\eea
In general, derivatives of $\tet$ functions do not transform well
under modular transformations. However, the following identity
overcomes this difficulty and leads to the expression
(\ref{Xi6})
announced earlier for $d\mu_2[\delta]$
\bea {\cal
M}_{\nu_1\nu_2}^2=\pi^4 \tet[\delta]^2\prod_{k=3,4,5}\tet
[\nu_1+\nu_2+\nu_k]^2.
\eea

\medskip

$\bullet$ To evaluate the $N$-point function, we need to evaluate
the contributions of the
vertex operators as well as of the component $d\mu_0[\delta]$ of the string chiral
measure. Since the relative phases $\epsilon_\delta$ of the GSO projection have
been already determined to be $1$, it suffices to consider the sum over spin structures
$\delta$ of these contributions with these phases.
In this case, clearly the split gauge is not appropriate since it is
$\delta$-dependent. Instead, we shall work in the unitary gauge,
where the points $q_\alpha$ are chosen to be the zeroes of a fixed holomorphic
$(1,0)$-form $\varpi(z)$
\bea
\varpi(q_1)=0,\quad \varpi(q_2)=0.
\eea
This gauge has the very important property that there exists a single-valued scalar
function $\Lambda(z)$ satisfying
\bea
\hat \mu_{\bar z} {}^z & = & S_\delta (q_1,q_2) \mu (z)
\no \\
\mu (z) \varpi(z) & = & \p_{\bar z}\Lambda(z).
\eea
We then need many $\tet$ function identities, of which the most difficult
are perhaps the ones involving the fermion stress tensor, and hence the term
\bea
\varphi[\delta](w;z_1,z_2)
=
S_\delta(z_1,w)\p_wS_\delta(w,z_2)
-
S_\delta(z_2,w)\p_wS_\delta(w,z_1).
\eea
For the $N$-point function with $N\leq 3$,
the existence of the function $\Lambda(z)$ turns out to imply the integral
identities
\bea
&\int \mu(w) I_{13}(w;z_1,z_2) =0
\nonumber\\
&\int \mu(w)\varpi(w)\{I_{14}(w;z_1,z_2,z_3)
+I_{14}(w;z_2,z_3,z_1)
+
I_{14}(w;z_3,z_1,z_2)\}=0
\eea
where the expressions $I_{13}(w;z_1,z_2)$ and $I_{14}(w;z_1,z_2,z_3)$
are defined by
\bea
I_{13}(w;z_1,z_2)&=&\sum_\delta{\cal Z}[\delta]S_\delta(q_1,q_2)\varphi[\delta](w;z_1,z_2)S_\delta(z_2,z_1)
\nonumber
\\
I_{14}(w;z_1,z_2,z_3)
&=&
\sum_\delta{\cal Z}[\delta]S_\delta(q_1,q_2)\varphi[\delta](w;z_1,z_2)S_\delta(z_2,z_3)S_\delta(z_3,z_1)
\eea
These identities imply in turn that $\sum_\delta {\cal B}[\delta]=0$ for $N\leq 3$.

\medskip

$\bullet$ The 4-point function is considerably more complicated, since we need to
extract Dolbeault exact differentials from $\sum_\delta {\cal B}[\delta]$ before
we can arrive at a holomorphic and gauge-independent form ${\cal H}$.
Also, we need identities of two types, those involving sums with ${\cal Z}[\delta]$,
and those involving sums with $\Xi_6[\delta]$. We illustrate these identities with
some examples. Consider first the sums involving ${\cal Z}[\delta]$
\bea
I_{15}(w;z_1,z_2,z_3,z_4)
&=&\sum_\delta {\cal Z}[\delta]\,S_\delta(q_1,q_2)\varphi[\delta](w;z_1,z_2)
S_\delta(z_2,z_3)S_\delta(z_3,z_4)S_\delta(z_4,z_1)
\nonumber\\
I_{16}(w;z_1,z_2,z_3,z_4)
&=&\sum_\delta {\cal Z}[\delta]\,S_\delta(q_1,q_2)\varphi[\delta](w;z_1,z_2)
S_\delta(z_2,z_1)S_\delta(z_3,z_4)^2,
\nonumber\\
I_{15}^S(w;z_1,z_2,z_3,z_4)
&=& {1\over 2}(I_{15}(w;z_1,z_2,z_3,z_4)+I_{15}(w;z_2,z_1,z_3,z_4))
\nonumber\\
I_{15}^A(w;z_1,z_2,z_3,z_4)
&=& {1\over 2}(I_{15}(w;z_1,z_2,z_3,z_4)-I_{15}(w;z_2,z_1,z_3,z_4)),
\eea
their integrated versions,
\bea
{\cal I}_{15}(z_1,z_2,z_3,z_4)
&=&{1\over 2\pi}\int \mu(w)I_{15}(w;z_1,z_2,z_3,z_4)
\nonumber\\
{\cal I}_{16}
&=&{1\over 2\pi}\int \mu(w)I_{15}(w;z_1,z_2,z_3,z_4).
\eea
the following cyclically permuted integrated versions,
\bea
{\cal I}_{15}^C(z_1,z_2,z_3,z_4)
&=&
+{\cal I}_{15}(z_1,z_2,z_3,z_4)
+
{\cal I}_{15}(z_2,z_3,z_4,z_1)
+
{\cal I}_{15}(z_3,z_4,z_1,z_2)
\nonumber\\
&&
+
{\cal I}_{15}(z_4,z_1,z_2,z_3)
\nonumber\\
{\cal I}_{16}^C (z_1,z_2;z_3,z_4)
&=&
+{\cal I}_{16}(z_1,z_2,z_3,z_4)
+
{\cal I}_{16}(z_3,z_4,z_1,z_2),
\eea
their symmetrized versions,
\bea
3\,{\cal I}_{15}^S(z_1,z_2,z_3,z_4)
&=&
{\cal I}_{15}^C(z_1,z_2,z_3,z_4)
+
{\cal I}_{15}^C(z_1,z_3,z_4,z_2)
+
{\cal I}_{15}^C(z_1,z_4,z_2,z_3)
\nonumber\\
3\,{\cal I}_{16}^S(z_1,z_2,z_3,z_4)
&=&
{\cal I}_{16}^C(z_1,z_2;z_3,z_4)
+
{\cal I}_{16}^C(z_1,z_3;z_4,z_2)
+
{\cal I}_{16}^C(z_1,z_4;z_2,z_3),
\eea
and their anti-symmetrized versions,
\bea
3\,{\cal I}_{15}^A(z_1,z_4|z_2,z_3)
&=&{\cal I}_{15}^C(z_1,z_2,z_3,z_4)-{\cal I}_{15}^C(z_1,z_3,z_2,z_4)
\nonumber\\
3\,{\cal I}_{16}^A(z_1,z_4|z_2,z_3)
&=&
{\cal I}_{16}^C(z_1,z_2;z_3,z_4)-{\cal I}_{16}^C(z_1,z_3;z_2,z_4).
\eea
Then we have the following identities
\bea
&{\cal I}_{15}^S(z_1,z_2,z_3,z_4)
=
-2{\cal I}_{16}^S(z_1,z_2,z_3,z_4)
=
-4{\cal Z}_0\,\sum_{i=1}^4\p\Lambda (z_i)
\prod_{j\not=i}\varpi(z_j)
\nonumber\\
&{\cal I}_{15}^A(z_1,z_4|z_2,z_3)
=
-{\cal I}_{16}^A(z_1,z_4|z_2,z_3)
=
{\zeta^1\zeta^2\over 4\pi^2}\Delta(z_1,z_4)\Delta(z_2,z_3),
\eea
where $\Delta(z,w)$ is the bi-holomorphic form in $z$, $w$ introduced earlier, and ${\cal Z}_0$
is the following quantity,
\bea
{\cal Z}_0
={Z^{12}\over
\pi^{12}\,\Psi_{10}(\Omega)\,E(q_1,q_2)^2\sigma(q_1)^2\sigma(q_2)^2},
\eea
with $Z$ the partition function of a single chiral boson,
expressible in terms of arbitrary points $r_1,r_2,r_3$,
\bea
Z^3
=
{\tet(r_1+r_2-r_3-\Delta)E(r_1,r_2)\sigma(r_1)\sigma(r_2)
\over
E(r_1,r_3)E(r_2,r_3)\sigma(r_3)\,{\rm det}\,\omega_I(r_j)}.
\eea

\medskip

Next, consider the sums involving $\Xi_6[\delta](\Omega)$
\bea
I_{20}(z_1,z_2;z_3,z_4)
&=&\sum_\delta \Xi_6[\delta]\tet[\delta]^4S_\delta(z_1,z_2)^2S_\delta(z_3,z_4)^2
\nonumber\\
I_{21}(z_1,z_2,z_3,z_4)
&=&
\sum_\delta \Xi_6[\delta]\tet[\delta]^4S_\delta(z_1,z_2)S_\delta(z_2,z_3)
S_\delta(z_3,z_4)S_\delta(z_4,z_1).
\eea
Then we have the identities
\bea
I_{20}(z_1,z_2;z_3,z_4)
&=&-4\pi^4\Psi_{10}(\Omega)(\Delta(z_1,z_3)\Delta(z_2,z_4)
+
\Delta(z_1,z_4)\Delta(z_2,z_3)
\nonumber\\
I_{21}(z_1,z_2,z_3,z_4)
&=&4\pi^4\Psi_{10}(\Omega)(\Delta(z_1,z_2)\Delta(z_3,z_4)
-
\Delta(z_1,z_4)\Delta(z_2,z_3)
\eea
as well as the identity
\bea
\sum_{IJKL}\omega_I(z_1)\omega_J(z_2)\omega_K(z_3)\omega_L(z_4)
\sum_\delta
\Xi_6[\delta]\tet[\delta]^3
\p_I\p_J\p_K\p_L\tet[\delta](0)=0.
\eea
All these identities combine to give the desired formulas for
$\sum_\delta{\cal B}[\delta](z_i;k_i,\epsilon_i;p_I^\mu)$
and ${\cal H}(z_i;k_i,\epsilon_i;p_I^\mu)$.
A crucial phenomenon is that all effects of gauge choices
reside only in the exact differentials ${\cal R}(z_i;k_i,\epsilon_i;p_I^\mu)$
which drops
out of the final physical amplitude ${\bf A}_{II}(k_i,\epsilon_i)$,
and that ${\cal H}(z_i;k_i,\epsilon_i;p_I^\mu)$
is completely gauge independent.

\subsection{Proof of non-renormalization theorems}

To obtain scattering amplitudes in the heterotic string, we combine
the anti-holomorphic factors from the type II
superstring amplitudes with the holomorphic factors from the 10-dimensional
bosonic string and internal fermions. The correlation functions
can be evaluated in a straightforward manner. The main issue
in the non-renormalization theorems is whether the
poles in the Mandelstam variables $s_{ij}=-2k_i\cdot k_j$
from the holomorphic sector survive after combination with
the anti-holomorphic sector and integration on the worldsheet.
The most difficult amplitudes are the $R^2F^2$ and the $R^4$ amplitudes,
so we discuss them briefly.

\medskip
For $R^2F^2$, the holomorphic sector is
${\cal W}={\cal W}_{(R^2)}(z_1,z_2){\cal W}_{(F^2)}(z_3,z_4)$,
with
\bea
{\cal W}_{(R^2)}=(\epsilon_1\cdot\epsilon_2)\,\p_{z_1}\p_{z_2}G(z_1,z_2)
-
\sum_{ij}(\epsilon\cdot k_i)\,(\epsilon\cdot k_j)\,\p_{z_1}G(z_1,z_j)
\p_{z_2}G(z_2,z_j)
\eea
This term leads to poles in $s_{ij}$. However, up to total derivatives on $\Sigma$,
$s_{ij}{\cal W}_{(R^2)}$ can be replaced by expressions of the form
\bea
s_{12}{\cal W}_{(R^2)}
\to
2(f_1f_2)\p_{z_1}\p_{z_2}G(z_1,z_2)-2
\sum_{ij}k_i^\mu f_1^{\mu\nu}f_2^{\nu\rho}k_j^\rho\p_{z_1}G(z_1,z_i)\p_{z_2}G(z_2,z_j)
\eea
Since the anti-holomorphic sector $\overline{{\cal Y}_S}$ always includes an $s_{ij}$ factor,
and since the above right hand side integrates to $0$ against anti-holomorphic forms,
these amplitudes do not contribute to the low-energy effective action.

\medskip
For $R^4$, the holomorphic sector is given by the expression ${\cal W}={\cal W}_{(R^4)}$
in (\ref{calW}). In this case, only the expressions $s_{ij}s_{kl}{\cal W}_{(R^4)}$
can be replaced, up to total
derivatives and to terms which vanish when $s_{ij}\to 0$, by sum of regular expressions
tending to $0$ as $s_{ij}\to 0$. However, when we expand the exponential factor
${\rm exp}(\sum_{ij}s_{ij}G(z_i,z_j))$ at low energy, we find that the contributions
of the constant term integrate to 0. Thus we need only consider the terms from
the exponential factor with at least one power of $s_{ij}$. Combined with the
other factor $s_{kl}$ from $\overline{{\cal Y}_S}$, we can apply then the
previous result for $s_{ij}s_{kl}{\cal W}_{(R^4)}$, and obtain the desired
non-renormalization theorem.

\subsection{S-duality and factorization}

The expressions (\ref{AII}) determine the superstring scattering amplitudes only up
to a constant factor depending only on the topology of the worldsheet. This
constant factor $C_2$ should be determined ultimately by the factorization
properties of the physical amplitudes. To compare with the S-duality
predictions for the two-loop correction to the $D^4R^4$ term in the effective
action, we need to compare two non-vanishing quantities, and the above constant factor
has to be determined precisely.
For this, we have to analyze the contributions in ${\bf A}_{II}(k_i,\epsilon_i)$
of the region of the moduli space ${\cal M}_2$ near the divisor of separating
nodes, and identify the resulting pole in $s\equiv s_{12}$
at $s={4\over\alpha'}$. Restoring the string tension parameter $\alpha'$,
the coupling constant $\lambda$, and the normalization $\kappa$
for the massless vertex operators,
we can write the amplitude ${\bf A}_{II}$ as
\bea
{\bf A}_{II}(k_i,\epsilon_i)
=
C_2
e^{2\lambda}K\bar K\kappa^4
\int_{{\cal M}_2}
{|\prod_{I\leq J}d\Omega_{IJ}|^2
\over
({\rm det\,Im}\,\Omega)^5}
\int_{\Sigma^4}|{\cal Y}_S|^2
{\rm exp}\bigg(-{\alpha'\over 2}\sum_{i<j}k_i\cdot k_jG(z_i,z_j)\bigg)
\eea
and we find \cite{dgp}
\bea
{\cal A}_{II}
=
-\delta(k){2^6\pi^3C_2/\alpha'
\over
s-4/\alpha'}e^{2\lambda}
K\bar K
{\cal B}_1^{(3)}(k_1,k_2,-q){\cal B}_1^{(3)}(k_3,k_4,q)
\eea
where ${\cal B}_1^{(3)}(k_1,k_2,q)$ and ${\cal B}_1^{(3)}(k_3,k_4,-q)$
are one-loop 3-point functions given by
\bea
{\cal B}_1^{(3)}(k_1,k_2,-q)
&=&
\int_{{\cal M}_1}{|d\tau_{11}|\over |Im\,\tau_{11}|^5}
\int d^2z_1d^2z_2{\rm exp}{\alpha's\over 4}\{G(z_1,z_2)-G(z_1,p_1)
-G(z_2,p_1)\}
\nonumber\\
{\cal B}_1^{(3)}(k_3,k_4,q)
&=&
\int_{{\cal M}_1}{|d\tau_{22}|\over |Im\,\tau_{22}|^5}
\int d^2z_3d^2z_4{\rm exp}{\alpha's\over 4}\{G(z_3,z_4)-G(z_3,p_2)
-G(z_4,p_2)\}
\nonumber
\eea
Comparing this with the factorization of tree-level and one-loop
amplitudes, we obtain the desired constant $C_2$. With the normalization
for the tree-level 4-point function given in \cite{dgp}, we have
$C_2={\sqrt 2\over 2^6(\alpha')^5}$.

Taking the limit $k_i\to 0$
in $A_{II}$, the low-energy
two-loop contribution to the $D^4R^4$ is then found to be
\bea
A_2^{(D^4R^4)}=8\,V_2C_2 e^{2\lambda}(\alpha')^2(s^2+t^2+u^2)\kappa^4K\bar K
\eea
where $V_2$ is the volume of the fundamental domain of $Sp(4,{\bf Z})/{\bf Z}_2$.
This volume has been determined by Siegel \cite{siegel}, and
combined with the value for $C_2$ just found, we find complete agreement with
S-duality.

\subsection{Orbifolds and KKS models}

The difficulties with gauge-fixing superstring amplitudes reside
only with the superghost part of the theory. For more general
space-times, the same method applies and gives well-behaved amplitudes,
as long as the earlier matter part $x^\mu$, $\psi_\pm^\mu$ is replaced
by a compactification which respects world-sheet supersymmetry.

\medskip
For ${\bf Z}_2$ orbifold models, the essential new features
are the twisted bosonic propagator $B_\epsilon(z,w)=\<\p_zx(w)\p_wx(w)\>_\epsilon$,
and the supersymmetric extension of the Prym period matrix.
The first is found to be
\bea
B_\epsilon(z,w)=S_{\delta_i^+}(z,w)S_{\delta_i^-}(z,w)
+b_i\omega_\epsilon(z)\omega_\epsilon(w)
\eea
where $\omega_\epsilon(z)$ is the Prym differential. On the other hand, there are
subtleties with the second: while the supersymmetric extension $\hat\Omega_{IJ}$
can be identified both as period matrix of a new complex structure $\hat g_{mn}$
and as covariance matrix of the chirally split amplitudes \cite{dp89},
the Prym matrix $\hat\tau_\epsilon$ of $\hat g_{mn}$ and the covariance $\tilde\tau_\epsilon$
of the chirally split twisted amplitudes are distinct supersymmetric
extensions of the Prym matrix $\tau_\epsilon$. Their difference is
\bea
\Delta\tau_\epsilon
=
-{i\over 8\pi}
\int\int d^2zd^2w\chiz S_\delta(z,w)\chi_{\bar w}{}^+ \left \{\omega_\epsilon(z)\omega_\epsilon(w)
-
\omega_I(z)\omega_J(w){\p\hat\tau_\epsilon\over\p\hat\Omega_{IJ}} \right \}.
\eea
where $\hat\tau$ is viewed as a function of
$\hat\Omega_{IJ}$. Only after taking properly into account such corrections can we arrive
at the correct ${\bf Z}_2$ gauge-fixed orbifold measure.

\newpage
\section{Directions for Further Investigation}
\setcounter{equation}{0}

In this section, we discuss a number of directions for possible further investigation.

\subsection{Higher genus superstrings}

The solution of two-loop superstrings gives us some optimism for
an eventual complete solution of superstring perturbation theory. Nevertheless,
the two-loop case benefits of a number of simplifying features: the $\bar\partial$
operator is always invertible for even spin structures $\delta$, the super period
matrix $\hat\Omega_{IJ}$ of a supergeometry is always defined
(instead of away from a subvariety),
and we can construct explicitly the fiber of supermoduli space over a fixed
$\hat\Omega_{IJ}$. In the bosonic string at 3-loops, it has been pointed out
that the spurious poles in the bosonic string integrand resulting
from the $\tet$ divisor can be cancelled by the zeroes from the measure
$\prod_{I\leq J}d\Omega_{IJ}$ \cite{perelomov}. We can hope that a similar mechanism
will take
place for the superstring. However, a manageable construction of the
fibers remains a challenging problem, and clearly much work will be needed.

\medskip
Alternatively, we can look for Ans\"atze for the 3-loop superstring
measure from factorization constraints, now that the 2-loop measure is known.
For example, if in analogy with the 2-loop case,
we take as Ans\"atz for the 3-loop string measure an
expression of the form,
\bea
d\mu[\Delta](\Omega^{(3)})
=
{\tet[\Delta](\Omega^{(3)})^4\Xi_6[\Delta](\Omega^{(3)})
\over
8\pi^4\Psi_9(\Omega^{(3)})}
\prod_{I\leq J}d\Omega_{IJ}^{(3)}
\eea
where $\Psi_9(\Omega^{(3)})^2=\prod_{\Delta\ even}\tet[\Delta](\Omega^{(3)})$ is Igusa's
modular form \cite{igusa}, then $\Xi_6[\Delta](\Omega^{(3)})$ must be
a modular covariant form of weight 6
satisfying the factorization constraint
\bea
{\rm lim}_{t\to 0}
\Xi_6[\Delta](\Omega^{(3)})
=
\eta(\Omega^{(1)})^{12}\,\Xi_6[\delta](\Omega^{(2)})
\eea
in the limit where the genus 3 surface with period matrices degenerates
into surfaces of genera $1$ and $2$ with period matrices $\Omega^{(1)}$
and $\Omega^{(2)}$, and the genus 3 spin structure $\Delta$
factors into two even spin structures. Polynomials in $\tet$ constants have been
found which can be candidates for $\Xi_6[\Delta](\Omega^{(3)})^2$.
It may be valuable to pursue this further \cite{dp04}.

\medskip
In another direction, we may try and generalize directly
the higher genus 4-point function from the very simple final expression (\ref{AII}) for
genus 2 and factorization properties. Several candidates have now been
proposed along these lines \cite{highergenus}.

\subsection{Odd spin structures}

For $N$ sufficiently large, the odd spin structures of the worldsheet
$\Sigma$ will begin contributing to the $N$-point function.
It would be important to extend our gauge-fixing method
to this case as well.
The chiral splitting of the matter fields $x^\mu$, $\psi_\pm^\mu$ has been carried
out in \cite{dp89} for odd spin structures $\delta$. A new
phenomenon is the emergence of an additional
superholomorphic form $\hat\omega_0$ which is the supersymmetric
extension of the holomorphic form $h_\delta(z)$ on $\kappa_\delta^{1/2}(\Sigma)$.
This is a source of new difficulties, since
from a certain point of view, the analogue of the super period matrix is now
$(h+1)\times (h+1)$ dimensional.

\subsection{BRST formalism}

The BRST symmetry is a powerful symmetry of gauge fixed quantum field theories
and particularly of string theories. Higher loop superstring amplitudes
based on BRST symmetry have been proposed a long time ago by Friedan,
Martinec, and Shenker \cite{fms}. However, the BRST invariance guarantees
the gauge slice independence of these amplitudes only up to total derivatives
on local patches on moduli space \cite{vv87}. In retrospect, we see that the
gauge-fixing method based on super period matrices has produced both local
and global corrections to the BRST prescription, under the form of an
insertion of the stress tensor and of the finite-dimensional determinants
in (\ref{calZgeneral}). It may be valuable to re-examine the amplitudes in this light,
and determine whether they can be arrived at by a BRST-like prescription.
The BRST formalism has also been re-examined for the bosonic string in
\cite{brst}, from other considerations.

\subsection{Effective actions}

As we had mentioned earlier, the two-loop amplitudes allow us to
determine the two-loop corrections to the effective action, and as
a by product, to get an indirect check of the many dualities
conjectured in string theory
\cite{Witten:1995ex, Bachas:1997xn, Tseytlin:1995fy}. The consistency with the conjectured
S-duality of the type IIB superstring has now been checked.
However, the relation with dualities of the non-renormalization
of the $\R^4$ term and the correction to the $D^2F^4$ term in the heterotic string
is still obscure (see \cite{dgp} and references therein).

\subsection{Normalizations of determinants}

The bosonization formulas of Fay \cite{fay}, Faltings \cite{faltings}, and
Verlinde-Verlinde \cite{vv87-2}
determine the chiral determinants of $\bar\partial$ operators up to constants depending
only on the genus. The exact value of these constants for the
$\bar\partial$ operator on scalars has received significant attention over the years
\cite{determinants}.
It would be useful to determine them for the $\bar\partial$
operator for all weights.

\end{document}